\documentclass[aps,reprint,twocolumn,superscriptaddress,longbibliography]{revtex4-1}
\usepackage{graphicx}
\usepackage{amssymb}
\usepackage{amsmath}
\usepackage{amsfonts}
\usepackage{bm}
\usepackage{braket}
\usepackage{xcolor}
\usepackage[colorlinks=True,citecolor=blue,urlcolor=blue,linkcolor=red]{hyperref}

\newcommand{\up}{\uparrow}
\newcommand{\dn}{\downarrow}

\newcommand{\s}{\mathrm{s}}
\renewcommand{\a}{\mathrm{d}}
\newcommand{\tar}{{\text{s}_\oplus}}
\newcommand{\aklt}{\text{AKLT}}
\newcommand{\tot}{\text{tot}}
\newcommand{\hsd}{H_\text{s-d}}
\newcommand\p[1]{\mathcal{P}_{S^\tot=2}^{(#1,#1+1)}}

\newcommand{\Pl}{\mathcal{P}}
\newcommand{\Plp}{\mathcal{P}^\prime}

\newcommand{\tr}{\mathrm{Tr}}

\newcommand\bovermat[2]{%
  \makebox[0pt][l]{$\smash{\overbrace{\phantom{%
    \begin{matrix}#2\end{matrix}}}^{\text{#1}}}$}#2}

\newcommand\partialphantom{\vphantom{A}}
\def\vecsign{\mathchar"017E}
\def\dvecsign{\smash{\stackon[-1.95pt]{\vecsign}{\rotatebox{180}{$\vecsign$}}}}
\def\dvec#1{\def\useanchorwidth{T}\stackon[-4.2pt]{#1}{\,\dvecsign}}
\usepackage{stackengine}
\stackMath

\begin{document}

\title{Measurement-induced steering of quantum systems}

\author{Sthitadhi Roy}
\email{sthitadhi.roy@chem.ox.ac.uk}
\affiliation{Rudolf Peierls Centre for Theoretical Physics, Clarendon Laboratory, Oxford University, Parks Road, Oxford OX1 3PU, United Kingdom}
\affiliation{Physical and Theoretical Chemistry, Oxford University, South Parks Road, Oxford OX1 3QZ, United Kingdom}

\author{J.~T.~Chalker}
\email{john.chalker@physics.ox.ac.uk}
\affiliation{Rudolf Peierls Centre for Theoretical Physics, Clarendon Laboratory, Oxford University, Parks Road, Oxford OX1 3PU, United Kingdom}

\author{I.~V.~Gornyi}
\email{igor.gornyi@kit.edu}
\affiliation{Institute for Quantum Materials and Technologies, Karlsruhe Institute of Technology, 76021 Karlsruhe, Germany}
\affiliation{A. F. Ioffe Physico-Technical Institute of the Russian Academy of Sciences, 194021 St. Petersburg,
Russia}

\author{Yuval Gefen}
\email{yuval.gefen@weizmann.ac.il}
\affiliation{Department of Condensed Matter Physics, Weizmann Institute of Science, 7610001 Rehovot, Israel}

\date{\today}

\begin{abstract}
We set out a general protocol for steering the state of a quantum system from an arbitrary initial state towards a chosen target state by coupling it to auxiliary quantum degrees of freedom. The protocol requires multiple repetitions of an elementary step: during each step the system evolves for a fixed time while coupled to auxiliary degrees of freedom (which we term `detector qubits') that have been prepared in a specified initial state. The detectors are discarded at the end of the step, or equivalently, their state is determined by a projective measurement with an unbiased average over all outcomes. The steering harnesses back-action of the detector qubits on the system, arising from entanglement generated during the coupled evolution. We establish principles for the design of the system-detector coupling that ensure steering of a desired form. We illustrate our general ideas using  both few-body examples (including a pair of spins-1/2 steered to the singlet state) and a many-body example (a spin-1 chain steered to the Affleck-Kennedy-Lieb-Tasaki state). We study the continuous time limit in our approach and discuss similarities to (and differences from) drive-and-dissipation protocols for quantum state engineering. Our protocols are amenable to implementations using present-day technology. Obvious extensions of our analysis include engineering of other many-body phases in one and higher spatial dimensions, adiabatic manipulations of the target states, and the incorporation of active error correction steps.
\end{abstract}
\maketitle

\tableofcontents

\section{Introduction \label{sec:introduction}}
There are many circumstances in which one would like to initialise a many-body quantum system in a specified state: examples of current interest range from quantum information processing to studies of non-equilibrium dynamics.  
Two standard approaches
for preparing quantum states are suggested by the laws of quantum mechanics and statistical mechanics. One of these is to make projective measurements of a set of observables represented by commuting operators that fully specify the target state. Alternatively, if the target state is the ground state of the Hamiltonian for the system, it can be reached by putting the system in thermal contact with a heat bath that is at a sufficiently low temperature. Both approaches have disadvantages, especially for a system with a large number of degrees of freedom: in the first approach, a general initial state is not an eigenstate of the measurement operator and hence the measurement outcome is probabilistic. The probability that the target state is reached decreases rapidly towards zero with increasing system size. With the second approach, the temperature scale required to completely eliminate thermal excitations from a system also decreases towards zero with size. 
In principle, if the initial state of the system is precisely known, a further possibility is to act on the state with an appropriately chosen perturbation for a precise interval of time, so that it evolves into the target state; this, however, requires extreme fine-tuning in a large system.

Our aim in the following is to establish a class of protocols for quantum state preparation that improve on both projective measurement and thermal contact with a heat bath. A protocol of the type we describe will, in an ideal implementation, steer a system from an arbitrary initial state to the target state, with guaranteed success. It does so by coupling the quantum system of interest to external detector qubits (auxiliary quantum degrees of freedom) that have been prepared in specified initial states, then evolving the coupled system under standard unitary quantum dynamics for a fixed time interval, and finally decoupling and discarding the detector qubits. 
Multiple repetitions of this process using freshly prepared detector qubits on each occasion, coupled to the system with a suitable interaction Hamiltonian, produce the outcome we require. In this paper we illustrate our general approach both for small systems consisting of one or two spin-1/2 degrees of freedom and for a macroscopic many-body system. 

This protocol and possible generalisations have the fundamental concepts of quantum entanglement and quantum measurements as essential ingredients, and possess conceptual links to the theory of open quantum systems.
First, the effect of a coupling between the system and detector qubits is to generate entanglement between their states. 
Second, focussing on the behaviour of the system after discarding the detector qubits, this entanglement induces steering of its quantum state, of a kind first discussed by Schr\"odinger in the context of the Einstein-Podolsky-Rosen (EPR) paradox~\cite{schrodinger1935discussion,schrodinger1936probability}.
A physical setting for the decoupling and discarding of the 
detectors
is that of performing strong measurements on these qubits, and then taking an unbiased average over the outcomes. 
Indeed, Schr\"odinger's original formulation of 
steering~\cite{schoedinger1929die} was in terms of a sophisticated experimenter performing suitable measurements on one of the parts of a bi-partite system to drive the other part to a state, chosen by the experimenter, with non-zero probability. More recently, the scope of the term “quantum steering” has been narrowed to the impossibility of local hidden state models describing the ensemble of states that a system can take upon measurement of another system entangled with it~\cite{wiseman2007steering,cavalcanti2016quantum,uola2019quantum}. We however continue to use the term in its broader sense.

To set out in more detail the links between our protocol for preparing or steering a quantum state and discussions of quantum measurement, it is useful to recall the distinction between \emph{strong} or \emph{projective measurements} and the notion of \emph{weak quantum measurements}~\cite{neumann1932mathematische,neumann2018mathematical,aharonov1988how,mello2010measurements,mello2014neumann,tamir2013introduction,svensson2013pedagogical}. The former
destroy the coherent quantum dynamics of the system. By contrast, under the latter, auxiliary degrees of freedom are coupled weakly to the system and projective measurements are performed on these detector qubits after decoupling them from the system.
No matter how weak the coupling, a measurement of this kind unavoidably impacts the system through its back-action, even after the detector qubits are decoupled from the system. 
Instead of viewing the measurement-induced disturbance on the system's state as a handicap, it may be considered a resource for quantum control and manipulation. The back-action of measuring the state of detectors entangled with the system can be harnessed to control the system's evolution and steer its state towards a desired target state: this constitutes the central message of our work.
In principle, one could make use of the measurement outcomes to determine whether the state of the system has been successfully steered in a realisation of the experiment, and discard it if this is not the case. In the simplest version of such \emph{postselected} protocols, the probability of success then goes down exponentially with the length of the outcome sequence or the number of detectors. This suppression is also likely to be exponentially strong in system size for a many-body system. To avoid such suppression, our aim is to establish protocols within a setting where the outcomes are averaged over in an unbiased fashion, and hence there is no loss of probability. We refer to processes of this type as \emph{blind measurements} or \emph{non-selective measurements}.

The concept of weak measurements allows for the notion of continuous measurements~\cite{brun2002simple,korotkov2003noisy,clerk2010introduction}. The latter can be viewed as a discrete sequence of weak measurements in the limit of vanishing time-interval for each weak measurement. This results in a sequence of measurement outcomes defining a quantum trajectory.
The equation for the density matrix of the system averaged over the trajectories is governed by a Lindblad equation.
In our case the steering protocol consists of a discrete sequence of measurements on detector qubits not necessarily coupled weakly to the system. A suitably defined time-continuum limit maps the dynamics of the density matrix of the system to a Lindblad equation.   For part of the analysis in our work, we exploit this formal connection between our steering protocol and Lindblad dynamics to gain insight into the steady state and the rate of approach to it.

The emergence of Lindblad dynamics suggests a comparison of
our protocol with proposals for preparing non-trivial many-body states in open quantum systems,
where the dissipative environment is posited to be Markovian and is treated within the framework of Lindblad dynamics~\cite{diehl2008quantum,kraus2008preparation,roncaglia2010pfaffian,weimer2010rydberg,lanyon2011universal,diehl2011topology,ticozzi2012stabilizing,bardyn2013topology,leghtas2013stabilizing,liu2016comparing}. The challenge in that context is to find suitable Lindblad jump operators that can be cast in terms of physical systems participating in the dynamical process. 
Our measurement-induced steering protocol has some technical similarities with this framework, as well as conceptual distinctions from it. With regard to the former, if the detector qubits we discuss are viewed as an environment, this environment is Markovian by construction, since the detector qubits are prepared afresh at the start of every cycle. Hence, the map governing the evolution of the density matrix of the system has a representation in terms of Kraus operators, which ultimately leads to a Lindblad equation in the time-continuum limit~\cite{lindblad1976generators,gorini1976completely}. 

An obvious advantage of our protocol is that the jump operators in the emergent Lindblad equation are uniquely and automatically fixed by the details of the system-detector coupling Hamiltonian, thus facilitating ``on-demand" engineering of Lindbladians. 
On the other hand, since our protocol is fundamentally a microscopic measurement protocol, certain requirements can easily be relaxed to go far beyond what is describable within the standard Lindblad framework. We note earlier works in the context of quantum control, discussing manipulation of quantum states via repeated interactions with designed environments~\cite{altafini2012modeling,ticozzi2012stabilizing,ticozzi2017alternating}; much of this, however, has focussed on the formal aspects of the theory or on applications restricted to few-body systems.

Another advantage of a measurement-based protocol over coupling with an environment is that, whereas the latter simply gets entangled with the system, a detector qubit can be read out, yielding information on the system state which could be further used to accelerate convergence to the desired state. Obvious examples include the use of post-selected protocols or using the measurement outcomes to implement active error correction 
and enhance the blind measurement protocols~\cite{belavkin1992quantum,minev2019catch}.

To summarise, our measurement-based steering protocol differs fundamentally from and has various advantages over related protocols for manipulating and stabilising non-trivial quantum states. First, the measurements in our case need not be weak: the detectors may be strongly coupled to the system. Second, in an appropriate time-continuum limit the dynamics is described by a Lindblad equation, but rather than introducing a Markovian reservoir or jump operators ``by hand", bath coupling is engineered by the measurement operators, which fix the jump operators uniquely. Third, unlike unstructured environments or baths, the detector qubits can be read out, with the outcomes exploited to further the efficiency and scope of our steering on platforms employing postselected protocols or using active decision making protocols based on the read-out sequence. Finally, we note the non-triviality of applying our measurement-based steering to many-body systems using an extensive number of auxiliary degrees of freedom.

\subsection*{Structure of the paper}

The rest of the paper is structured as follows. We start with an overview in Sec.~\ref{sec:overview},
where we introduce the basic ingredients of the steering protocol and state the main results of the work.
Section~\ref{sec:formalism} presents the details of the steering protocol. 
The guiding principle for designing
the protocol is introduced and derived in Sec.~\ref{sec:map}, followed by the derivation of the effective Lindblad dynamics in Sec.~\ref{sec:lindblad}. The workings and advantage of the approach is exemplified via the simple case of a spin-1/2 pair steered towards their singlet state in Sec.~\ref{sec:spin1/2}. We then turn to a many-body quantum system, a spin-1 chain -- which we steer to the 
Affleck-Kennedy-Lieb-Tasaki (AKLT) ground state ~\cite{affleck1987rigorous,affleck1988valence} 
in Sec.~\ref{sec:spin1}. As a building block of the protocol, the steering of a spin-1 pair is discussed in Sec.~\ref{sec:pairspin1}, followed by the numerical treatment of the spin-1 chain in Sec.~\ref{sec:aklt}. We close with discussions of possible experimental implementations of the protocol and future directions in Sec.~\ref{sec:discussion}.

\section{Overview \label{sec:overview}}
The central result of this work is a general formalism for 
measurement-induced 
steering of a many-body quantum system towards a non-trivial target state, by repeatedly coupling and decoupling a set of auxiliary degrees of freedom interspersed with unitary dynamics of the composite system.
The protocol makes use of the entanglement generated between the system and the auxiliary degrees of freedom to steer the state.
The auxiliary degrees of freedom are simple quantum systems with small Hilbert-space dimensions such that they are easy to prepare in a desired initial state; in this work we consider a set of decoupled detector qubits (spins-1/2) initially polarised along a given direction.

\subsection{Steering protocol}
The steering protocol can be described as a sequence of discrete steering events each of which consists of the following steps:
\begin{itemize}
	\item[] The detector qubits are prepared in a initial given state, which does not depend on the state of the system. We denote the initial state of the detector qubits as $\ket{\Phi_\a}$ and the density matrix corresponding to this initial state as $\rho_\a$.
	\item[] The system is then coupled to the detector qubits and the composite system evolves unitarily under a Hamiltonian for some time. Denoting the density matrix of the system at time $t$ as $\rho_\s(t)$ and the system-detector Hamiltonian as $H_{\s\text{-}\a}$, the state of the system-detector composite after evolution for an interval $\delta t$ is given by
	\begin{equation}
		\rho_{\s\text{-}\a}(t+\delta t) = e^{-i \hsd\delta t}\rho_\a\otimes\rho_\s(t)e^{i \hsd\delta t}.
		\label{eq:evol}
	\end{equation}
	\item[] The detector qubits are then decoupled from the system.
	 Formally we may trace out the detector degrees of freedom to obtain the density matrix of the system at time $t+\delta t$
	\begin{equation}
		\rho_\s(t+\delta t) = \mathrm{Tr}_\a \rho_{\s\text{-}\a}(t+\delta t).
		\label{eq:map}
	\end{equation}
	Equations~\eqref{eq:evol} and \eqref{eq:map} describe the discrete time-evolution map for the density matrix of the system under the steering dynamics.
	\item[] The detector qubits are re-prepared in their initial states and the above steps are repeated.
\end{itemize}

\subsection{Steering inequalities}

As we would like the system to get steered towards the target state, denoted by $\ket{\Psi_\tar}$ (with corresponding density matrix $\rho_\tar$), the dynamics induced by $\hsd$ should satisfy the steering inequality
\begin{equation}
	\braket{\Psi_\tar|\rho_\s(t+\delta t)|\Psi_\tar} \ge \braket{\Psi_\tar|\rho_\s(t)|\Psi_\tar},\,\forall~t,
	\label{eq:golden-inequality}
\end{equation}
with the inequality ideally becoming an equality only if $\rho_\s(t)= \rho_\tar$.

Note that this inequality is very strong. When combined with the condition on equality, they ensure that the system is steered to the target state irrespective of its initial state.
In principle, one could envisage weaker steering inequalities. One example is
\begin{equation}
\lim_{t\to\infty}\braket{\Psi_\tar|\rho_\s(t)|\Psi_\tar}=1
\end{equation} and a still weaker one is
\begin{equation}
\lim_{t\to\infty}\braket{\Psi_\tar|\rho_\s(t)|\Psi_\tar} > \braket{\Psi_\tar|\rho_\s(0)|\Psi_\tar}\,.
\end{equation}
We present in Sec.~\ref{sec:map} a general strategy for designing system-detector couplings so that the strongest of these forms, \eqref{eq:golden-inequality}, holds.

\subsection{Guiding principle for choice of system-detector coupling Hamiltonian \label{sec:guiding}}
With the protocol fixed, it remains to find a guiding principle for the choice of the Hamiltonian that governs the evolution of the system-detectors composite.
As a first step, consider summing both side of \eqref{eq:golden-inequality} over $\rho_\s(t)$ representing all possible pure states. If \eqref{eq:golden-inequality} is obeyed, then $W \equiv e^{-i \hsd\delta t}$ must satisfy
\begin{equation}
{\rm Tr}_{\rm s,d} [W\rho_\a\otimes\openone_\s W^\dagger\cdot\openone_\a\otimes\rho_\tar ] \geq {\rm Tr}_{\rm s,d} [\rho_\a \otimes \rho_\tar]\,.
\end{equation}
In the special case where the Hilbert spaces of the system ($\mathcal{H}_\s$) and the detector ($\mathcal{H}_\a$) have the same dimension, we see from this that for the inequality to be as strong as possible, $W$ should be the swap operator between these two spaces and $\rho_\a$ should be the image of $\rho_\tar$ under this swap. More generally,
a Hamiltonian which contains direct products of operators $O_\a^{(n)}$ in the detectors' subspace $\mathcal{H}_\a$ that rotate the detectors from their initial state to an orthogonal subspace and operators $U_\s^{(n)}$ in the system's subspace $\mathcal{H}_\s$ that rotate the system to the target state manifold from an orthogonal subspace, will steer the system towards the target state.
Such a Hamiltonian has the form
\begin{equation}
	\hsd = \sum_{n}\left(O_\a^{(n)}\ket{\Phi_\a}\bra{\Phi_\a}\right)\otimes U_\s^{(n)} + \mathrm{h.c.},
	\label{eq:hamcoupling}
\end{equation}
where $n$ labels the detector qubit.
Since $O_\a^{(n)}$ connects the state $\ket{\Phi_\a}$ to its orthogonal subspace, it satisfies $\braket{\Phi_\a|O^{(n)}_\a|\Phi_\a}=0$. Additionally, we assert the following properties for the system operators: 
\begin{itemize}
\item [(i)]$U_\s^{(n)}$ annihilates the target state, $U_\s^{(n)}\ket{\Psi_\tar}=0$
\item [(ii)] $U_\s^{(n)}$ is normalised such that $U_\s^{(n)}U_\s^{(n)\dagger}\ket{\Psi_\tar}=\ket{\Psi_\tar}$ 
\item [(iii)] for different $n$ and $m$, if $U_\s^{(n)}$ and $U_\s^{(m)}$ share spatial support then $U_\s^{(m)}U_\s^{(n)\dagger}=0$. If they do not share a spatial support, they automatically commute.
\end{itemize}
To ensure that equality in Eq.~(\ref{eq:golden-inequality}) always implies $\rho_\s(t) = \rho_\tar$ it is necessary that $|\Psi_\tar\rangle$ is coupled to every state in its complement by at least one operator $U_s^{(n)\dagger}$. In our study of steering to the AKLT state, we observe guaranteed convergence to the target state despite the set of operators $\{U_s^{(n)\dagger}\}$ not connecting the AKLT state to all the other states in the Hilbert space. However, monotonicity is not ensured and, in principle, can depend on the particular measure of distance from the target state. We disucss this issue in detail in Secs.~\ref{sec:map}, \ref{sec:aklt} and Appendix~\ref{sec:aklt-monotonic}.

\subsection{Toy example with a spin-1/2 \label{subsec:toy}}
As a simple example consider a single spin-1/2 which we wish to steer to the fully polarised state along the $z$-direction, so that $\ket{\Psi_\tar}=\ket{\up_\s}$. 
The subspace orthogonal to $\ket{\Psi_\tar}$ has only a single state: $U_\s^\dagger\ket{\Psi_\tar}=\ket{\dn_\s}$. 
Hence, a single detector qubit is sufficient to steer the state. Without loss of generality, let us choose it to be initialised as $\ket{\Phi_\a}=\ket{\up_\a}$.
Naturally, we also have $O_\a\ket{\Phi_\a}=\ket{\dn_\a}$.
Following Eq.~\eqref{eq:hamcoupling}, the system-detector coupling Hamiltonian can then be written as 
\begin{equation}
	\begin{aligned}
		\hsd &= J (\ket{\dn_\a}\bra{\up_\a})\otimes(\ket{\up_\s}\bra{\dn_\s}) + \mathrm{h.c.} \\
		  &= J\left(\sigma^-_\a\sigma^+_\s + \sigma^+_\a\sigma^-_\s\right),
	\end{aligned}
	\label{eq:ham-single-spin}
\end{equation}
where $\sigma_{\s(\a)}^\pm$ denote the Pauli raising and lowering operators for the system and detector qubits.

At time $t$, the state of the system can be generally written as $\rho_\s(t) = (\mathbb{I}_2+\mathbf{s}(t)\cdot\bm{\sigma}_\s)/2$ where $\mathbf{s}(t)$ is a classical three-component vector. The state of the system spin-1/2 at time $t+\delta t$ can then be obtained by evolving the combined state of the system and the detector, $(\mathbb{I}_2+\sigma^z_\a)\otimes\rho_\s(t)/2$, with the unitary operator $W$, and tracing over the detector.
This yields the relation for $\mathbf{s}(t+\delta t)$
\begin{equation}
	\begin{aligned}
		s_x(t+\delta t) &= \cos(J\delta t)s_x(t) = \cos^{\frac{t}{\delta t}}(J\delta t)s_x(0),\\
		s_y(t+\delta t ) &= \cos(J\delta t)s_y(t)= \cos^{\frac{t}{\delta t}}(J\delta t)s_y(0),\\
		s_z(t+\delta t) &= 1 - \cos^2(J\delta t) + \cos^2(J\delta t)s_z(t)\\
						&= 1 - \cos^\frac{2t}{\delta t}(J\delta t) + \cos^\frac{2t}{\delta t}(J\delta t)s_z(0).
	\end{aligned}
	\label{eq:single-spin-components-recursion}
\end{equation}
The above equation explicitly shows that
\begin{equation}
	\lim_{t\to\infty}\mathbf{s}(t) = (0,0,1),
	\label{eq:single-spin-infinite-time}
\end{equation}
irrespective of the initial conditions and that the limit is approached exponentially in time.

This example illustrates how
 a system-detector coupling Hamiltonian of the form in Eq.~\eqref{eq:hamcoupling}  leads to the satisfaction of the desired steering inequality, Eq.~\eqref{eq:golden-inequality}, for generic initial states of the system. This guarantees successful steering to the target state from arbitrary initial states. In turn, the example shows how weak measurements can produce control and manipulation that projective measurements cannot~\footnote{It is clear that a projective measurement of $\sigma^z$ will not guarantee steering to $\ket{\up}$ as the measurement would project the spin to either $\ket{\up}$ or $\ket{\dn}$ with probabilities $\mathrm{Tr}[\rho_\s(t) (1\pm\sigma^z)/2]$ respectively, $\rho_\s(t)$ being the state of the system spin-1/2.}.

We remark that although in the above example the target state of the system qubit is the same as the initial state of the detector qubit, there is nothing special about this choice: see Appendix~\ref{sec:arbitrary}.

\subsection{Many-body states and multiple detectors}
In the context of many-body systems, the subspace orthogonal to the target is exponentially large in the system size. Hence it may appear that the system-detector coupling Hamiltonian of the form in Eq.~\eqref{eq:hamcoupling} entails an exponentially large number of detectors, or non-local system-detector coupling Hamiltonians, or both. For implementation of the protocol to be feasible, we would like to have at most an extensive number of detectors, with couplings that are local. 
Although these constraints appear rather restrictive, they can be satisfied for versions of our protocol that
allow steering to a large class of strongly correlated quantum states which are eigenstates of unfrustrated local projector Hamiltonians. One can steer to such states by locally steering different parts of the system to the appropriate eigenstate of the corresponding local projector part of the Hamiltonian.
Since the effective Hilbert dimension of a local part of the system is finite, one can steer locally with a finite number of detectors, and hence the total number of detectors required scales linearly with system size, and not exponentially. Moreover, the couplings are manifestly local.

The particular example we consider in detail is the Affleck-Kennedy-Lieb-Tasaki (AKLT) state of a one-dimensional spin-1 chain~\cite{affleck1987rigorous,affleck1988valence}. The AKLT state is a valence bond state such that on each bond between two neighbouring spins-1, there is no projection on the total spin-2 sector. Thus one can steer the spin-1 chain by locally steering each bond out of the total spin-2 sector, and the uniqueness of the ground state of the AKLT chain (with periodic boundary conditions) guarantees steering to the AKLT state. Steering each bond requires only a finite number of detectors due to the finite dimension of the total spin-2 subspace for a pair of spins-1 and hence the total number of detectors needed to steer a chain is only extensive in system size.

Further examples of states which satisfy the above criteria include matrix product states, projected pair entangled states, the Laughlin state of a fractional quantum Hall system~\cite{haldane1983fractional}, the ground state of Kitaev's toric code~\cite{kitaev2003fault}, and of recent interest, fermionic symmetry-protected topologically ordered eigenstates of full commuting projector Hamiltonians~\cite{son2018commuting,tantivasadakarn2018full}.

An obvious potential concern arises when our protocol is used to steer many-body systems to eigenstates of local projector Hamiltonians, because in the cases of interest a given quantum degree of freedom appears in more than one projector. 
In particular, in such a scenario it is not guaranteed that two system operators $U_\s^{(n)}$ and $U_\s^{(m)}$ acting on such a shared degree of freedom, while steering their corresponding parts of the system locally, satisfy the conditions listed at the end of Sec.~\ref{sec:guiding}.
Steering towards the target space of one projector may therefore undo, at least partially, the effect of steering to the target space of a different projector with which the degree of freedom in question is shared.
Despite these complications, we show that our approach enables us to select zero-energy eigenstates of local projector Hamiltonians as target states and unique stationary states of our dynamics. We illustrate this numerically for the AKLT chain, showing steering to the ground state with guaranteed success from arbitrary initial states.

\subsection{Steady states and rate of approach}

Equations~\eqref{eq:evol} and \eqref{eq:map} show that the time-evolution of the density matrix of the system is governed by a linear map. 
On general grounds, the largest eigenvalue magnitude for this map is unity. 
If the associated eigenvector is unique,
this ensures that there exists a unique steady state. The logarithm of the next eigenvalue then encodes the rate of approach in time to this steady state. Ideally, we would like the rate to be finite in the thermodynamic limit so that the time-scale required to steer the many-body state arbitrarily close to the target state does not diverge with system size.

In the context of the AKLT state, we find that 
the magnitude of these eigenvalues is most conveniently studied 
in the time-continuum limit ($\delta t\to 0$) of the map in Eq.~\eqref{eq:map} where an effective Lindblad equation emerges for the equation of motion of $\rho_\s$ with the jump operators being fixed by the system-detectors coupling Hamiltonian $\hsd$. 
Unit eigenvalue for the discrete map corresponds to zero eigenvalue for the Lindbladian operator.
If the corresponding Lindbladian eigenvector is unique, it is the stationary state of the steering protocol.
Successful steering implies that the steady state of the Lindbladian is the target state.
Moreover, the gap in the spectrum of the Lindbladian between the zero eigenvalue and the rest of the eigenvalues determines the rate of the of approach of the system to the target state.
In the case of the AKLT state, analysis  of the size-dependence of the numerically obtained gap in the effective Lindbladian's spectrum  indicates that it stays finite in the thermodynamic limit.

\section{Formalism of quantum state steering  \label{sec:formalism}}
	In this section, we describe the quantum steering protocol in detail. 
	We show explicitly that a system detector coupling Hamiltonian of the form given in Eq.~\eqref{eq:hamcoupling} leads to dynamics that satisfy the steering inequality, Eq.~\eqref{eq:golden-inequality}, 
	on rather general grounds.
	Additionally, we make the connection to an effective Lindblad equation that describes the time-continuum limit of the steering map, Eq.~\eqref{eq:map}.

	\subsection{General derivation of steering inequality from discrete-time map \label{sec:map}}
	Let us discuss in some generality the map governing the discrete time evolution of the system density matrix, Eq.~\eqref{eq:map}, with the system-detector coupling Hamiltonian $\hsd$ of Eq.~\eqref{eq:hamcoupling}.
	We will first show that $\rho_\s=\rho_\tar=\ket{\Psi_\tar}\bra{\Psi_\tar}$ is indeed a stationary state of the map and then show that the system is progressively steered towards the $\rho_\tar$ after each steering event.

	That $\rho_\tar$ is a stationary state of the map is easily shown by noting that with $\hsd$ from Eq.~\eqref{eq:hamcoupling}, $\hsd \cdot (\rho_\a\otimes\rho_\tar) = 0$, which naturally implies
	\begin{equation}
		e^{-i \hsd\delta t}\rho_\a\otimes\rho_\tar e^{i \hsd\delta t} = \rho_\a\otimes\rho_\tar.
		\label{eq:steadystate}
	\end{equation}
	Upon taking the trace over the detectors, one recovers that $\mathrm{Tr}_\a [e^{-i \hsd\delta t}\rho_\a\otimes\rho_\tar e^{i \hsd\delta t}] = \rho_\tar$ and hence $\rho_\tar$ is a stationary state of the map.
	
	We next show that the system is steered towards the target state at every discrete steering event. 
	For brevity, we show here the derivation with a single detector qubit and state the results for multiple detector qubits. Details of the derivation for the latter are presented in Appendix.~\ref{sec:derivation-multiple}.

	To proceed, we find it most convenient to partition the composite Hilbert space of the detectors and the system, $\mathcal{H}_\a\otimes \mathcal{H}_\s$ into two subspaces, which we denote as $\mathcal{D}_\a\otimes\mathcal{H}_\s$ and $\overline{\mathcal{D}}_\a\otimes\mathcal{H}_\s$ where $\mathcal{D}_\a$ is the one-dimensional subspace spanned by $\ket{\Phi_\a}$ and $\mathcal{D}_\a\oplus\overline{\mathcal{D}}_\a = \mathcal{H}_\a$.
	In this convention, the density matrix $\rho_\a\otimes\rho_\s(t)$ can be represented as
	\vspace{0.25cm}
	\begin{equation}
		\begin{matrix}
 			\rho_\a\otimes\rho_\s(t)
 			=
 			\begin{bmatrix}
 				\bovermat{$\mathcal{D}_\a$}{~~\rho_\s(t)~~} & \bovermat{$\overline{\mathcal{D}}_\a$}{~~~~0~~~}\\
 				~~0~~ & ~~~~~0~~~~\\
  			\end{bmatrix}
  			\begin{aligned}
  				&\left.\begin{matrix}
  				\partialphantom 
  				\end{matrix} \right\} %
  				\mathcal{D}_\a\\
  				&\left.\begin{matrix}
  				\partialphantom
  				\end{matrix}\right\}%
  				\overline{\mathcal{D}}_\a\\
 				\end{aligned}
 		\end{matrix}.
 		\label{eq:rho-matrix-bipartition}
 	\end{equation}
 	Additionally, the system-detector coupling Hamiltonian, Eq.~\eqref{eq:hamcoupling} takes the form
 	\vspace{0.25cm}
 	\begin{equation}
		\begin{matrix}
 			\hsd
 			=
 			\begin{bmatrix}
 				\bovermat{$\mathcal{D}_\a$}{~~~0~~~} & \bovermat{$\overline{\mathcal{D}}_\a$}{~~~U^\dagger~~~}\\
 				~~~U~~~ & ~~~0~~~\\
  			\end{bmatrix}
  			\begin{aligned}
  				&\left.\begin{matrix}
  				\partialphantom 
  				\end{matrix} \right\} %
  				\mathcal{D}_\a\\
  				&\left.\begin{matrix}
  				\partialphantom
  				\end{matrix}\right\}%
  				\overline{\mathcal{D}}_\a\\
 				\end{aligned}
 		\end{matrix}.
 	\end{equation}
 	The unitary time-evolution operator generated by the system-detector coupling Hamiltonian is
 	\begin{equation}
		e^{-i\hsd\delta t}
		=
		\sum_{k=0}^\infty\frac{(-i\delta t)^{2k}}{(2k)!}\left[\begin{smallmatrix}
			\left(U^\dagger U\right)^k &\left(\frac{-i\delta t}{2k+1}\right)U^\dagger\left(U U^\dagger\right)^k\\
			\left(\frac{-i\delta t}{2k+1}\right)U\left(U^\dagger U\right)^k &\left(U U^\dagger\right)^k\\
		\end{smallmatrix}\right].
  		\label{eq:unitary-full}	
 	\end{equation}
 	Following the steering map of Eq.~\eqref{eq:map}, applying the above unitary matrix to the density matrix of Eq.~\eqref{eq:rho-matrix-bipartition} and taking the trace of the detector, one obtains the time evolved density matrix of the system as
 	\begin{widetext}
 	\begin{align}
 		\rho_\s(t+\delta t) = &\sum_{k,l}\frac{(-i\delta t)^{2k}(i\delta t)^{2l}}{(2k)!(2l)!}\left[\left(U^\dagger U\right)^k\rho_\s(t)\left(U^\dagger U\right)^l\right]+
 		\sum_{k,l}\frac{(-i\delta t)^{2k+1}(i\delta t)^{2l+1}}{(2k+1)!(2l+1)!}\left[U\left(U^\dagger U\right)^k\rho_\s(t)U^\dagger\left(U U^\dagger\right)^l\right].
 		\label{eq:time-evolved-exact}
 	\end{align}
 	\end{widetext}
 	Using ${U_\s\ket{\Psi_\tar}}{=}0$ and $U_\s U_\s^\dagger\ket{\Psi_\tar}{=}\ket{\Psi_\tar}$, the diagonal matrix element of the time-evolved density matrix corresponding to the target state can be obtained from Eq.~\eqref{eq:time-evolved-exact} as
 	\begin{widetext}
 	\begin{equation}
 		\braket{\Psi_\tar|\rho_\s(t+\delta t)|\Psi_\tar}=\braket{\Psi_\tar|\rho_\s(t)|\Psi_\tar}+\underbrace{\braket{\Psi_\tar|U_\s\rho_\s(t)U_\s^\dagger|\Psi_\tar}}_{Q}\sin^2(\delta t).
 		\label{eq:diagelem-time-evolved-exact}
 	\end{equation}
 	\end{widetext}
 	Note $Q$ is a diagonal matrix element of a valid density matrix $\rho_\s(t)$ which ensures that it is non-negative.
 	The change in the diagonal element of the density matrix corresponding to the target state is proportional to the support of the density matrix on the subspace orthogonal to $\ket{\Psi_\tar}$.
 	Hence, the inequality in Eq.~\eqref{eq:golden-inequality} becomes an equality when $\rho_\s(t)$ has no remaining support on the subspace orthogonal to $\ket{\Psi_\tar}$ -- in other words, $\rho_\s(t)$ has reached its target form.
 	This concludes the derivation of the steering inequality, Eq.~\eqref{eq:golden-inequality}, which in turn shows that the system is steered towards the target state progressively at each steering event.

 	In the case of multiple detectors, the equation for the time-evolved density matrix has the form as in Eq.~\eqref{eq:diagelem-time-evolved-exact} but $Q$ is given by (see Appendix~\ref{sec:derivation-multiple} for derivation)
 	\begin{equation}
 		Q = \sum_{n}\braket{\Psi_\tar|U_\s^{(n)}\rho_\s(t)U_\s^{(n)\dagger}|\Psi_\tar}%\sin^2(\delta t).
 		\label{eq:Q-multiple}
 	\end{equation}
 	which is again a sum of diagonal elements of a density matrix and hence is non-negative. Also it is the total support of the density matrix on the subspace orthogonal to $\ket{\Psi_\tar}$ spanned by the set of states $\{U_\s^{(n)\dagger}\ket{\Psi_\tar}\}$. Provided $|\Psi_\tar\rangle$ is connected to every state in its complement by an operator $U_s^{(n)\dagger}$ then $Q=0$ implies $\rho_\s(t) = \rho_\tar$.

For a many-body system, if a sufficiently large number of operators is employed that the target state is connected to all others by the operators $U_s^{(n)\dagger}$, monotonic convergence to the target state is straightforwardly guaranteed. The required number of operators may, however, be exponentially large in system size. By contrast, we are interested in steering a many-body system using only an extensive number of operators, each coupled locally to the system. It can then happen that $Q=0$ for some $\rho_\s(t)$ at a given step, but that further steps successfully steer the system to the target. In Sec.~\ref{sec:aklt} we show how in a such a setting with extensive, local steering operators a spin-1 chain is steered uniquely to the AKLT state. In this case we also examine whether or not steering is monotonic. We show that this may depend on the choices of initial state and of the measure of the distance between the time-evolving state and the target state; see Appendix~\ref{sec:aklt-monotonic} for details. We will quantify the distance of $\rho_\s(t)$ from the target state, $\rho_\tar$, via the Frobenius and trace norms, denoted as $D_F$ and $D_1$ respectively. The are defined as
	\begin{subequations}
		\begin{align}
		\begin{split}
			D_F(t) =& \sqrt{\mathrm{Tr}[\rho_\s(t)-\rho_\tar]^2},
		\end{split}\\
		\begin{split}
			D_1(t) =& \mathrm{Tr}\left[\sqrt{(\rho_\s(t)-\rho_\tar)^2}\right]/2.
		\end{split}
		\end{align}
		\label{eq:matrix-norms}
	\end{subequations}
Additionally, since the AKLT ground state is an unique zero-energy ground state of the AKLT Hamiltonian (with periodic boundary conditions), another good measure for the distance is the energy of the state measured with respect to the AKLT Hamiltonian
\begin{equation}
E_\aklt(t) = \mathrm{Tr}[H_\aklt \rho_\s(t)].
\label{eq:EAKLT}
\end{equation}

	\subsection{Effective Lindblad dynamics \label{sec:lindblad}}

	We now show that the time-continuum limit of the map in Eq.~\eqref{eq:map} leads to a Lindblad equation for the dynamics of the density matrix of the system.
	The exposition of the conceptual connection between measurement-induced quantum steering and the dynamics being described by the Lindblad equation, often associated with dissipative dynamics, is an important result of this work.
	The Lindblad equation is a master equation for the density matrix of the form
	\begin{align}
		\partial_t\rho_\s  &= \mathcal{L}[\rho_\s]\nonumber\\&=-i[H_\s,\rho_\s ]+ \sum_i\left[L_i\rho_\s L_i^\dagger -\frac{1}{2}\{L_i^\dagger L_i,\rho_\s \}\right],
		\label{eq:lindblad}
	\end{align}
	where $H_\s$ is the intrinsic Hamiltonian of the system and the $L_i$s are the \emph{quantum jump operators}.
	They are fixed in our derivation by the form of the system-detector coupling Hamiltonian.
	
	As in Sec.~\ref{sec:map}, we sketch the derivation with a single detector and state the result for multiple detectors.
	To proceed with the derivation we again partition the composite Hilbert space and represent $\rho_\a\otimes\rho_\s(t)$ as in \eqref{eq:rho-matrix-bipartition}.
 	The general system-detector coupling Hamiltonian can be represented by a matrix of the form
 	\vspace{0.3cm}
 	\begin{equation}
		\begin{matrix}
 			\hsd
 			=
 			\begin{bmatrix}
 				\bovermat{$\mathcal{D}_\a$}{~~~V~~~} & \bovermat{$\overline{\mathcal{D}}_\a$}{~~~U^\dagger~~~}\\
 				~~~U~~~ & ~~~V^\prime~~~\\
  			\end{bmatrix}
  			\begin{aligned}
  				&\left.\begin{matrix}
  				\partialphantom 
  				\end{matrix} \right\} %
  				\mathcal{D}_\a\\
  				&\left.\begin{matrix}
  				\partialphantom
  				\end{matrix}\right\}%
  				\overline{\mathcal{D}}_\a\\
 				\end{aligned}
 		\end{matrix}.
 		\label{eq:H-matrix-bipartition}
 	\end{equation}
 	Expanding the map in Eq.~\eqref{eq:map} to second order in $\delta t$
	one obtains
 	\begin{align}
 		\frac{\rho_\s(t+\delta t)-\rho_\s(t)}{\delta t} =& i[V,\rho_\s(t)]-\frac{1}{2}[V,[V,\rho_\s(t)]]\delta t +\nonumber\\
 		&\left(U\rho_\s(t)U^\dagger-\frac{1}{2}\{U^\dagger U,\rho_\s(t)\}\right)\delta t.
 		\label{eq:evolution-second-order-general}
 	\end{align}
 	Taking the limit $\delta t\to 0$ with $\tilde{U}=U\sqrt{\delta t}$ while requiring $||V||\sim\mathcal{O}(1)$ and $||\tilde{U}||\sim\mathcal{O}(1)$~\footnote{Note that the $\sqrt{\delta t}$ in the rescaling of the energy-scales associated to the operator $U$ is strongly reminiscent of the scaling of the white noise term in an It\^{o} stochastic differential equation~\cite{vankampen1992stochastic}. This is not surprising as a stochastic Schr{\"o}dinger equation with white noise, when averaged over the noise, yields a Lindblad equation for the density matrix with the jump operators given by the terms to which the noise couples~\cite{cai2013algebraic}.},
 	one obtains a time-continuum equation of motion 
 	\begin{equation}
 		\partial_t\rho_\s(t) = -i[V,\rho_\s(t) ]+ \left[\tilde{U}\rho_\s(t) \tilde{U}^\dagger -\frac{1}{2}\{\tilde{U}^\dagger \tilde{U},\rho_\s(t) \}\right].
 		\label{eq:lindblad-derived}
 	\end{equation}
	Comparing with Eq.~\eqref{eq:lindblad}, it is trivial to see that the equation of motion for $\rho_\s$ is indeed of the Lindblad form with the jump operator given by $\tilde{U}$.
	The general Hamiltonian in Eq.~\eqref{eq:H-matrix-bipartition} reduces to the specific one in Eq.~\eqref{eq:hamcoupling} with the choice of operators $V=V^\prime=0$.
	Hence the effective Lindblad equation which describes the dynamics of the steering protocol in the time-continuum limit consists only of the jump operators.
	In the case of multiple detectors, the Lindblad equation obtained has multiple jump operators
 	\begin{equation}
 		\partial_t\rho_\s = -i[V,\rho_\s(t) ]+ \sum_n\left[\tilde{U}_n\rho_\s(t) \tilde{U}_n^\dagger -\frac{1}{2}\{\tilde{U}_n^\dagger \tilde{U}_n,\rho_\s(t) \}\right].
 		\label{eq:lindblad-derived-multiple}
 	\end{equation}

	A few comments regarding the Lindblad equation are in order which will eventually prove useful in the analysis of the concrete examples we discuss.
	Note that the Lindblad equation is linear in $\rho_\s$.
	Hence, if the $\mathcal{D}_{\mathcal{H}_\s}$-dimensional density matrix $\rho_\s(t)$ is unravelled as a supervector, $\vec{\rho_\s}(t)$, of dimension $\mathcal{D}_{\mathcal{H}_\s}^2$, the time-evolution is simply generated by the superoperator corresponding to the Lindbladian
	\begin{equation}
		\vec{\rho_\s}(t) = \exp\left(\dvec{\mathcal{L}}t\right)\cdot\vec{\rho_\s}(0).
		\label{eq:superoperator}
	\end{equation}
	The steady state manifold of the dynamics is then given by the manifold of states corresponding to zero eigenvalues of $\dvec{\mathcal{L}}$.
	Hence the uniqueness of the steady state can be determined by studying the degeneracy of the zero eigenvalues. 
	Moreover, the gap in the spectrum of $\dvec{\mathcal{L}}$ between the zero eigenvalue and the rest of the eigenvalues, $\{v\}$,defined as
	% ~\footnote{The eigenvalues are guaranteed to have negative real parts.}, 
		$\Delta = \min_{v\neq 0}\{\vert \mathrm{Re}[v]|\}$
	determines the nature of the approach to the steady state.
	For a finite $\Delta$ in the thermodynamic limit, the system approaches the steady state exponentially fast in time with a rate $\Delta$.

	In some of the examples we study, the Lindblad equation that arises has a special form.
	Specifically, if there exists a choice of basis $\{\ket{\alpha}\}$ such that all the jump operators are of the form $\sqrt{\gamma_{\alpha\beta}}\ket{\alpha}\bra{\beta}$, then the matrix elements of the density matrix in the same basis follow a master equation given by
	\begin{equation}
		\partial_t\rho_{\s,\alpha\beta} = \delta_{\alpha\beta}\sum_\nu\gamma_{\alpha\nu}\rho_{\s,\nu\nu} - \frac{1}{2}\rho_{\s,\alpha\beta}\sum_\nu(\gamma_{\nu\alpha}+\gamma_{\nu\beta}),
		\label{eq:lindblad-components}
	\end{equation}
	which directly implies that the off-diagonal elements of the density matrix decay exponentially in time. Moreover, the equations for the diagonal elements do not involve the off-diagonal elements and hence follow a classical master equation.

\subsection{Steering a pair of spins-1/2 \label{sec:spin1/2}}

	\begin{figure}
		\includegraphics[width=\columnwidth]{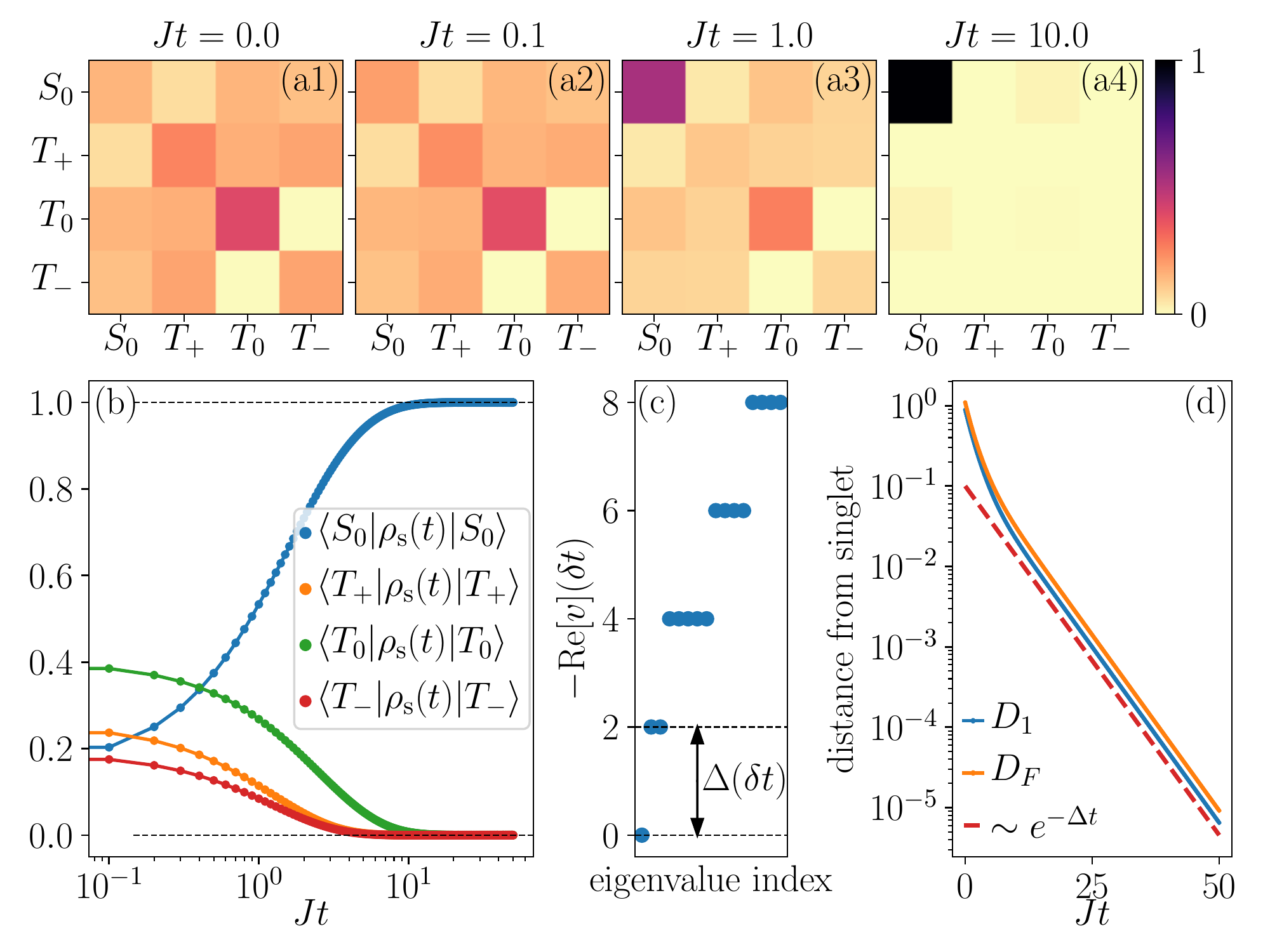}
		\caption{{\bf Steering of a pair of spins-1/2 to the singlet state} using three detector qubits coupled via the Hamiltonian in Eq.~\eqref{eq:hamcoupling-singlet}. (a) The density matrix of the system, $\rho_\s$ in the basis spanned by the singlet and the three triplets as a colourmap (for the absolute values of the matrix elements) at different times $t$. It shows the decaying support on the $S^\tot=2$ diagonal elements. (b) The evolution of the diagonal elements of $\rho_\s(t)$ shows the steering to the singlet. (c) The spectrum of the corresponding Lindbladian with a unique zero eigenvalue (steady state) and a finite gap $\Delta$. (d) The distance of  $\rho_\s(t)$ from the singlet state $\ket{S_0}\bra{S_0}$ as measured via the matrix norms in Eq.~\eqref{eq:matrix-norms} decays exponentially in time at a rate is correctly given by $\Delta$. For the plots, $J=1$ and $\delta t=0.1$.}
		\label{fig:spin-half-singlet}
	\end{figure}

	We next illustrate these ideas using a two-spin problem.
	Consider a pair of spins-1/2, acted on by the set of Pauli matrices, $\{\sigma^\mu_1\}$ and $\{\sigma^\mu_2\}$, which we would like to steer to the singlet state $\ket{S_0} = (\ket{\up\dn}-\ket{\dn\up})/\sqrt{2}$.
	The Hilbert space of two spins-1/2 is four-dimensional with the subspace orthogonal to $\ket{S_0}$ spanned by the three triplet states, $\ket{T_+} = \ket{\up\up}$, $\ket{T_-}=\ket{\dn\dn}$, and $\ket{T_0} = (\ket{\up\dn}+\ket{\dn\up})/\sqrt{2}$.
	Hence, the simplest steering protocol uses three detector qubits, each for steering the state out of one of the triplet states onto the singlet state.
	At the start of every steering event, each of the detector qubits is prepared in a pure state fully polarised along the positive $z$-axis such that
	\begin{equation}
		\rho_\a = \frac{\mathbb{I}_2+\sigma^z_{\a_1}}{2}\otimes\frac{\mathbb{I}_2+\sigma^z_{\a_2}}{2}\otimes\frac{\mathbb{I}_2+\sigma^z_{\a_3}}{2},
		\label{eq:detector-intialise-singlet}
	\end{equation}
	where $\{\sigma_{\a_i}^\mu\}$ denotes the set of Pauli matrices for the $i^{\mathrm{th}}$ detector.
	For this choice of $\rho_\a$, the system-detector coupling Hamiltonian motivated from the form in Eq.~\eqref{eq:hamcoupling} can be written as
	\begin{equation}
		\hsd = J\sum_{i=1}^3 \left(\sigma^-_{\a_i}\rho_\a\otimes U_i  + \mathrm{h.c.}\right),
		\label{eq:hamcoupling-singlet}
	\end{equation}
	with
	\begin{subequations}
		\begin{align}
		\begin{split}
		U_1 =\ket{S_0}\bra{T_+}= \frac{1}{2^{\frac{3}{2}}}[(1+\sigma^z_1)\sigma^-_2 - \sigma^-_1(1+\sigma^z_2)],
		\end{split}\\
		\begin{split}
		U_2 =\ket{S_0}\bra{T_-}= \frac{1}{2^{\frac{3}{2}}}[(1-\sigma^z_1)\sigma^+_2 - \sigma^+_1(1-\sigma^z_2)],
		\end{split}\\
		\begin{split}
		U_3 =\ket{S_0}\bra{T_0}= \frac{1}{2}\left[\sigma^+_1\sigma^-_2 - \sigma^-_1\sigma^+_2 + \frac{\sigma^z_1+\sigma^z_2}{2}\right].
		\end{split}
		\end{align}
		\label{eq:hamcoupling-singlet-terms}
	\end{subequations}
	\noindent 
	Note from the above equations that $U_iU_j^\dagger = 0$ for $i\neq j$ as each of the triplet states are orthogonal to each other and also to the target singlet state, and hence the conditions asserted for the system operators in Sec.~\ref{sec:guiding} are satisfied.
	Results for the evolution of the two-spin system with the above steering protocol are shown in Fig.~\ref{fig:spin-half-singlet}.
	All the elements of density matrix in the basis spanned by the singlet and the three triplet states decay to zero exponentially in time except for the diagonal element corresponding to the singlet, which approaches unity.
	This shows that the system is indeed steered to the singlet state exponentially in time.

	Turning to the effective Lindblad dynamics corresponding to the protocol, the three jump operators of the Lindblad equation can be read off from Eq.~\eqref{eq:hamcoupling-singlet} as $L_i = JU_i\sqrt{\delta t}$.
	The spectrum of the so-obtained Lindbladian superoperator, $\dvec{\mathcal{L}}$, is shown in Fig.~\ref{fig:spin-half-singlet}(c). 
	The zero eigenvalue is non-degenerate implying that the singlet state is the unique steady state of the dynamics. 
	Moreover, the distance of $\rho_\s(t)$ from the singlet state obtained from the exact dynamics and measured via the Frobenius or the trace norm (Eq.~\eqref{eq:matrix-norms}) decays exponentially in time with a rate given by $\Delta \times\delta t$ which is in excellent agreement to the numerically obtained gap in the spectrum of the Lindbladian between the zero eigenvalue and the rest of the eigenvalues.
	This shows that the effective Lindblad equation is a valid description of the dynamics.

	Note further that the form of the jump operators in Eq.~\eqref{eq:hamcoupling-singlet-terms} leads the equation of the density matrix elements in the basis spanned by the singlet and the triplets to be of the form in Eq.~\eqref{eq:lindblad-components}.
	This explains the exponential decay of the off-diagonal elements.
	Additionally, the master equation for the diagonal elements possesses only the loss term for the triplet states and only the gain terms for the singlet, resulting the exponential decay and growth of the former and latter respectively.

	Before concluding the section, it is worth mentioning that a protocol for steering a pair of spins-1/2 to the singlet using only a single detector qubit can also be envisaged. 
	Such a protocol involves steering the system out of any one of the triplet states onto the singlet state while also acting on it with an intrinsic Hamiltonian which has matrix elements connecting the other two triplet states to the first one.
	Consequently, the Hamiltonian unitarily rotates the states within the triplet sector and weight from the triplet sector keeps leaking into the singlet, eventually leading to the pure singlet state as the steady state. 
	For details, see Appendix~\ref{sec:shake-and-steer}.

\section{Spin-1 chain \label{sec:spin1}}
	
	Let us now discuss the steering of a quantum many-body system, namely that of the spin-1 chain to the AKLT ground state.
	The AKLT Hamiltonian is 
	\begin{equation}
		H_\aklt = \sum_{\ell=1}^N\left[\frac{1}{3}+\frac{1}{2}\hat{\mathbf{S}}_\ell\cdot\hat{\mathbf{S}}_{\ell+1}+\frac{1}{6}(\hat{\mathbf{S}}_\ell\cdot\hat{\mathbf{S}}_{\ell+1})^2\right],
		\label{eq:ham-aklt}
	\end{equation}
	where $\hat{\mathbf{S}}_\ell$ denotes spin-1 operators at site $\ell$ and the site $\ell=N+1$ is identified with $\ell=1$ to impose periodic boundary conditions.

	It is useful to set up the notation for the rest of the section here.
	The projectors onto the three eigenstates of $\hat{S}^z_\ell$ (with eigenvalues $\pm 1$ and 0 respectively), $P_\ell^\pm$ and $P_\ell^0$, are given by
	\begin{equation}
		P_\ell^\pm = [(S^z_\ell)^2\pm S^z_\ell]/2;~~P^0_\ell=\mathbb{I}_3-P_\ell^+ - P_\ell^-.
		\label{eq:projector-spin1}
	\end{equation}
	The total spin on a bond between the sites $\ell$ and $\ell+1$ will be denoted by $\mathbf{S}_{(\ell,\ell+1)}^\tot=\mathbf{S}_\ell+\mathbf{S}_{\ell+1}$. 
	In what follows, it will be often convenient to use the simultaneous eigenstates of $\left(\hat{\mathbf{S}}_{(\ell,\ell+1)}^\tot\right)^2$ and $\left(\hat{S}^{\tot,z}_{(\ell,\ell+1)}\right)$, denoted as $\ket{S^\tot,S^{\tot,z}}_{(\ell,\ell+1)}$ where $S^\tot$ takes values 0, 1, and 2, and $S^{\tot,z}$ takes all integer values in the range $-S^\tot\le S^{\tot,z}\le S^\tot$. Additionally, we denote the AKLT ground state as $\ket{\Psi_\aklt}$ and the corresponding density matrix as $\rho_\aklt = \ket{\Psi_\aklt}\bra{\Psi_\aklt}$.

	The AKLT Hamiltonian with periodic boundaries has a unique valence-bond ground state which is most easily understood by noting that $H_\aklt$ can be expressed as a sum of local projectors
	\begin{equation}
		H_\aklt = \sum_{\ell=1}^N  \mathcal{P}_{S^\tot=2}^{(\ell,\ell+1)},
		\label{eq:ham-aklt-projector}
	\end{equation}
	where $\mathcal{P}_{S^\tot=2}^{(\ell,\ell+1)}$ is a projector onto the five-dimensional $S^\tot=2$ subspace for the pair of spins-1 at sites $\ell$ and $\ell+1$.
	The form of $H_\aklt$ in Eq.~\eqref{eq:ham-aklt-projector} implies that the ground state 
	has the special property that each bond has zero weight in the $S^\tot=2$ sector.
	Hence, it is natural to imagine that a spin-1 chain can be steered to the AKLT ground state by locally steering each bond out of the $S^\tot=2$ sector.
	Moreover, since the $S^\tot=2$ sector of each bond is finite-dimensional, such a steering protocol would satisfy our physically motivated constraints: locality and only an extensive number of detectors.

	 In  Sec.~\ref{sec:pairspin1} we discuss the basic building block of the steering protocol, namely steering a pair of spins-1 out of the $S^\tot=2$ sector. 
	 In Sec.~\ref{sec:aklt} we use this building block to steer a spin-1 chain to the AKLT state.

	\subsection{A pair of spins-1 \label{sec:pairspin1}}

		Consider steering on the bond between a pair of spins-1 labelled as $\ell$ and $\ell+1$. 
		The $S^{\tot}_{(\ell,\ell+1)}=2$ subspace is five-dimensional and hence the simplest protocol involves five detector qubits, one to steer out of each of the $S^{\tot}_{(\ell,\ell+1)}=2$ states.
		Note that we only wish to steer out of the $S^{\tot}_{(\ell,\ell+1)}=2$ subspace and there are no restrictions on what precise state(s) in the $S^{\tot}_{(\ell,\ell+1)}=1$ and $S^{\tot}_{(\ell,\ell+1)}=0$ subspaces we steer onto.
		This offers a lot of freedom in choosing the steering protocol. This could be exploited to construct the simplest system-detector coupling Hamiltonian, or from a practical point of view, the one that it is easiest to implement.
		In the following, we choose a particular protocol which steers the state onto the $S^{\tot}_{(\ell,\ell+1)}=1$ subspace keeping the sign of $S^{\tot,z}_{(\ell,\ell+1)}$ the same.

		At the start of each steering event, the detector qubits~\footnote{Note that while the system now is composed of spins-1, we continue to use spins-1/2 as detectors.} are prepared in a polarised pure state
		\begin{equation}
			\rho_\a^{(\ell,\ell+1)} = \otimes\prod_{i=1}^5\frac{\mathbb{I}_2+\sigma^z_{\a_i^{(\ell,\ell+1)}}}{2},
			\label{eq:detector-intialise-spin1}
		\end{equation}
		and the system-detector coupling Hamiltonian is of the form
		\begin{equation}
		\begin{aligned}
			\hsd^{(\ell,\ell+1)} = J\sum_{i=1}^5 &\sigma^-_{\a_i^{(\ell,\ell+1)}}\rho_\a^{(\ell,\ell+1)}\otimes U_i^{(\ell,\ell+1)}+ \mathrm{h.c.},
			\label{eq:hamcoupling-spin1-pair}
		\end{aligned}
		\end{equation}
		where
		\begin{widetext}
			\begin{subequations}
			\begin{align}
			\begin{split}
			U_1^{(\ell,\ell+1)} =& \left(\ket{1,1}\bra{2,2}\right)_{(\ell,\ell+1)} = \frac{1}{2}[(S_\ell^- - S_{\ell+1}^-)P_\ell^+P_{\ell+1}^+],
			\end{split}\\
			\begin{split}
			U_2^{(\ell,\ell+1)} =& \left(\ket{1,1}\bra{2,1}\right)_{(\ell,\ell+1)} = \frac{1}{2}\left[\left(\mathbb{I}_9-\frac{S_\ell^-S_{\ell+1}^+}{2}\right)P_\ell^+P_{\ell+1}^0 - \left(\mathbb{I}_9-\frac{S_\ell^+S_{\ell+1}^-}{2}\right)P_\ell^0P_{\ell+1}^+\right],
			\end{split}\\
			\begin{split}
			U_3^{(\ell,\ell+1)} =& \left(\ket{1,0}\bra{2,0}\right)_{(\ell,\ell+1)} = \frac{1}{2\sqrt{3}}\left[\left(\mathbb{I}_9-\frac{(S_\ell^-S_{\ell+1}^+)^2}{4}\right)P_\ell^+P_{\ell+1}^- - \left(\mathbb{I}_9-\frac{(S_\ell^+S_{\ell+1}^-)^2}{4}\right)P_\ell^-P_{\ell+1}^+ \right.\nonumber
			\end{split}\\
			\begin{split}
						&~~~~~~~~~~~~~~~~~~~~~~~~~~~~~~~~~~~~~~\left.+ (S_\ell^+S_{\ell+1}^--S_\ell^-S_{\ell+1}^+)P_\ell^0P_{\ell+1}^0\right],
			\end{split}\\
			\begin{split}
			U_4^{(\ell,\ell+1)} =& \left(\ket{1,-1}\bra{2,-1}\right)_{(\ell,\ell+1)} = \frac{1}{2}\left[\left(\mathbb{I}_9-\frac{S_\ell^+S_{\ell+1}^-}{2}\right)P_\ell^-P_{\ell+1}^0 - \left(\mathbb{I}_9-\frac{S_\ell^-S_{\ell+1}^+}{2}\right)P_\ell^0P_{\ell+1}^-\right],
			\end{split}\\
			\begin{split}
			U_5^{(\ell,\ell+1)} =& \left(\ket{1,-1}\bra{2,-2}\right)_{(\ell,\ell+1)} = \frac{1}{2}[(S_\ell^+ - S_{\ell+1}^+)P_\ell^-P_{\ell+1}^-].
			\end{split}
			\end{align}
			\label{eq:spin1ops-steer-onto-stot=1}
			\end{subequations}
		\end{widetext}
		As in Sec.~\ref{sec:formalism}, the jump operators for the corresponding Lindblad equation in the time-continuum limit can be identified straightforwardly as 
		\begin{equation}
			L_i^{(\ell,\ell+1)} = J U_i^{(\ell,\ell+1)}\sqrt{\delta t}.
			\label{eq:jumpops-spin-1}
		\end{equation}

		\begin{figure}
			\includegraphics[width=\columnwidth]{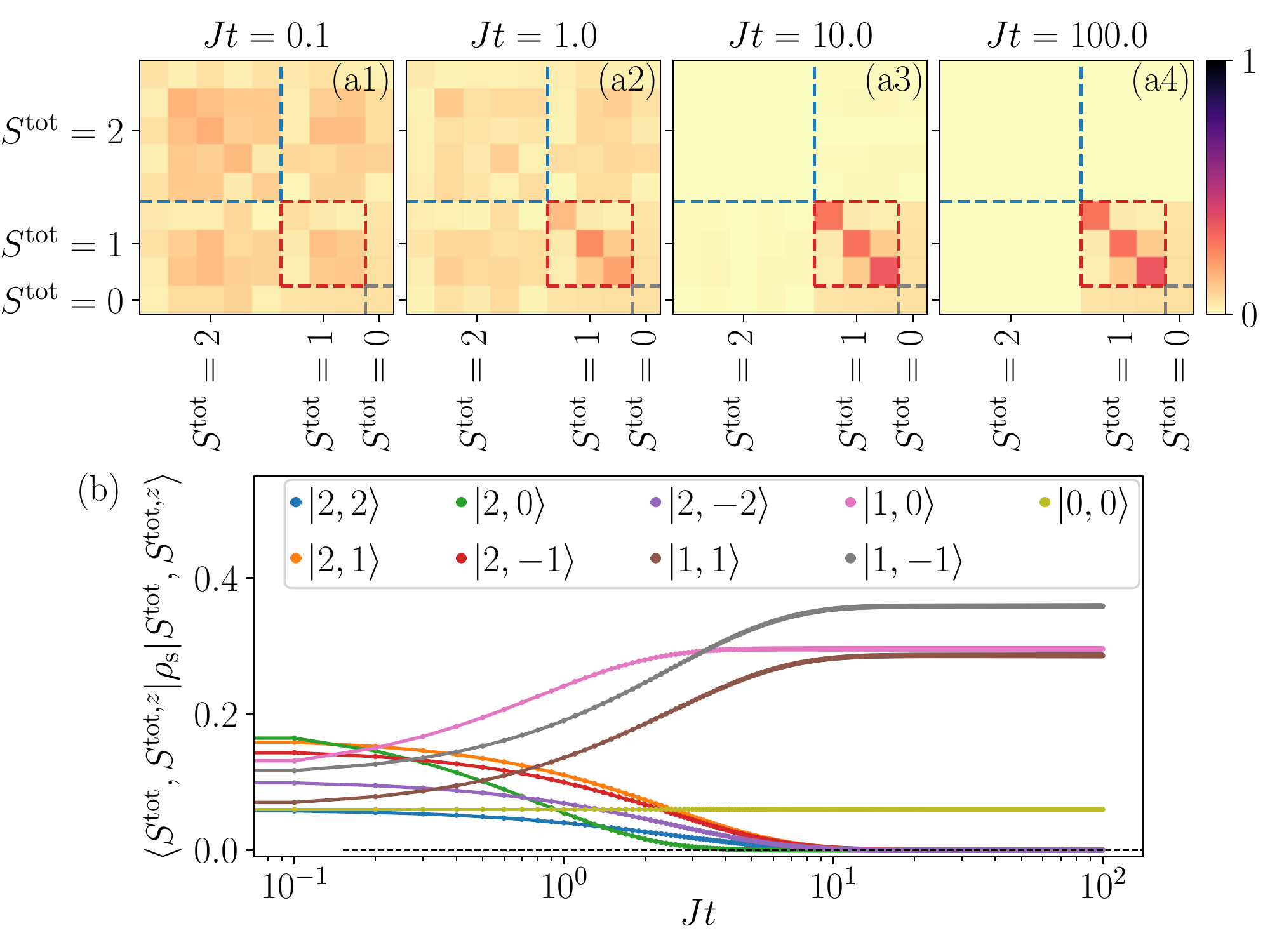}
			\caption{{\bf Steering of a pair of spins-1 out of the $\bm{S^\tot=2}$ subspace} using the system-detector coupling Hamiltonian of Eq.~\eqref{eq:hamcoupling-spin1-pair}. (a) The density matrix of the system, $\rho_\s$ in the $\ket{S^\tot,S^{\tot,z}}$ basis as a colourmap (for the absolute values of the matrix elements) at different times $t$: it shows the decaying and growing support on the $S^\tot=2$ and the $S^\tot=0$ and 1 subspaces respectively. The blocks demarcated by the blue, red, and gray dashed lines respectively denote the $S^\tot=2$, $1$, and $0$ sectors. (b) The evolution of the diagonal elements of $\rho_\s(t)$ shows the steering out of the $S^\tot=2$ subspace, all the five diagonal elements of which decay to zero. For the plots, $J=1$ and $\delta t=0.1$.}
			\label{fig:spin-1-pair}
		\end{figure}
		
		In the rest of this subsection, we only discuss the pair of spins-1 and hence drop the labels $\ell$ and $\ell+1$.
		We also refer to the density matrix of the two spins as $\rho_\s(t)$.
		The dynamics of $\rho_\s(t)$ under the steering protocol with the system-detector coupling Hamiltonian in Eq.~\eqref{eq:hamcoupling-spin1-pair} starting from a random mixed state is shown in Fig.~\ref{fig:spin-1-pair}.
		Expressed in the $\ket{S^\tot,S^{\tot,z}}$ basis, the matrix elements of $\rho_\s(t)$ in the $S^\tot=2$ block and the off-diagonal blocks connected to it decay exponentially in time.
		At long enough times, $\rho_\s(t)$ is supported entirely on the subspace spanned by the $S^\tot=1$ and $S^\tot=0$ states, and thus is steered out of the $S^\tot=2$ subspace.
		With regard to the corresponding Lindblad equation, note that the jump operators, Eq.~\eqref{eq:jumpops-spin-1}, for the system-detector coupling Hamiltonian, Eq.~\eqref{eq:spin1ops-steer-onto-stot=1}, result in an equation of motion of the form of Eq.~\eqref{eq:lindblad-components} for the matrix elements of $\rho_\s(t)$ in the $\ket{S^\tot,S^{\tot,z}}$ basis.
		Several features of the dynamics can be inferred from this observation.
		For any state $\ket{\alpha}$ in the $S^\tot=1$ or $S^\tot=0$ sector, $\gamma_{\nu\alpha}=0$ for $\nu\neq\alpha$ which renders the off-diagonal matrix elements within the $S^\tot=1$ and $S^\tot=0$ subspace invariant in time (see also Fig.~\ref{fig:spin-1-pair}(a1)-(a4)).
		This also implies that the master equations for the diagonal elements within this subspace have no loss terms and only gain terms, contrary to those in the $S^\tot=2$ subspace, which only have loss terms. 
		One can then immediately infer that the support of $\rho_\s(t)$ on the $S^\tot=2$ subspace decays exponentially, whereas it grows on the  $S^\tot=1$ and $S^\tot=0$ subspace before saturating to unity.

		An important point to notice is for a pair of spins-1, the above dynamics does not have a unique steady state unlike the case exemplified in Sec.~\ref{sec:spin1/2}.
		This is due to the fact that in the case of the spins-1, the steering is not onto a particular pure state but rather onto a subspace of states, and the specific steady state of the dynamics depends on the initial condition of the system.
		The non-uniqueness of the steady state for the pair of spins-1 is also manifested in the corresponding Lindbladian having a degenerate zero-eigenvalue manifold.
		In the case of a spin-1 chain, as we will discuss in the next subsection, the uniqueness of the AKLT ground state leads to the uniqueness of the steady state reached under dynamics obtained by extending the approach described above to all the bonds of the chain.

	\subsection{Steering to the AKLT state \label{sec:aklt}}

		Let us now discuss in detail the many-body case of a chain of spins-1.
		The protocol for steering the system to the AKLT ground state is a scaled-up version of the protocol described in the previous subsection where each bond is steered out of the $S^\tot=2$ subspace and as such there are $5N$ detector qubits at play, $N$ being the size of the spin-1 chain.
		Formally, the initial state of the detectors at the start of each steering event is expressed as
		\begin{equation}
			\rho_\a = \otimes\prod_{\ell=1}^N \prod_{i=1}^5\frac{\mathbb{I}_2+\sigma^z_{\a_i^{(\ell,\ell+1)}}}{2},
			\label{eq:detector-intialise-aklt}
		\end{equation}
		and the system-detector coupling Hamiltonian is of the form
		\begin{equation}
			\hsd = J\sum_{\ell=1}^N\sum_{i=1}^5 \left(\sigma^-_{\a_i^{(\ell,\ell+1)}}\rho_\a\otimes U_i^{(\ell,\ell+1)}\right)  + \mathrm{h.c.},
			\label{eq:hamcoupling-aklt}
		\end{equation}
		where $U_i^{(\ell,\ell+1)}$ is the same as in Eq.~\eqref{eq:spin1ops-steer-onto-stot=1}.

		Let us first argue that the protocol described by Eqs.~\eqref{eq:detector-intialise-aklt} and \eqref{eq:hamcoupling-aklt} does posses the AKLT ground state at least as one of its steady states.
		Note that each term in the system-detector coupling Hamiltonian in Eq.~\eqref{eq:hamcoupling-aklt} is of the form $\sigma^-_\a \otimes \ket{1,s_1}\bra{2,s_2}+ \mathrm{h.c.}$.
		Hence a state of the form $\rho_\a\otimes\rho_\aklt$ is annihilated by the Hamiltonian as $\sigma^+_\a\rho_\a = 0 = \ket{1,s_1}\bra{2,s_2}\rho_\aklt$, implying that $\rho_\aklt$ is indeed a steady state of the dynamics.
		However, compared to the `ideal' steering protocol described in Sec.~\ref{sec:formalism}, a number of difficulties arise for many-body systems. The difficulties arise because the Hilbert space for a many-body system is exponentially large in system size. The protocol of Sec.~\ref{sec:formalism} uses a separate term in $\hsd$ to steer to the target state from each other state. Such an approach is not practical for a many-body system, where one requires a limit of at most an extensive number of terms in $\hsd$. In the following we discuss these issues, and present numerical results for the AKLT chain which show successful steering in this many-body setting.

		In the derivation of the strong inequality for steering in the presence of multiple detectors (see Sec.~\ref{sec:guiding} and Appendix~\ref{sec:derivation-multiple}), we had required that $U_\s^{(m)}U_\s^{(n)\dagger}=0$ if $U_\s^{(m)}$ and $U_\s^{(n)}$ have overlapping spatial supports. For the system-detector couplings with Eq.~\eqref{eq:spin1ops-steer-onto-stot=1} for the AKLT chain, it can be shown that $U_i^{(\ell,\ell+1)} U_j^{(r,r+1)\dagger} \neq0$ if $\vert \ell-r\vert=1$. In other words, the assumption made in the derivation is not satisfied by the system-detector couplings on adjacent bonds but it is satisfied by all the other pairs of couplings. The non-commutativity of the steering operators on adjacent bonds may also lead to non-monotonic time-dependence for $E_\aklt(t)$. Steering a bond, say between sites $\ell$ and $\ell+1$, affects the state of three bonds: the one that is steered and the two adjacent to it. While the energy of the bond being steered necessarily goes down as steering decreases $\mathrm{Tr}\left[\p{\ell}\rho_\s(t)\right]$, the same need to be the case for $\mathrm{Tr}\left[\p{\ell^\prime}\rho_\s(t)\right]$ with $\ell^\prime=\ell\pm1$. For this reason, the total energy of the state need not decrease.

		The difficulty associated with the non-commutativity of the steering operators can be partially removed by considering a protocol in which each steering step is divided into multiple steps, in one of which only the system-detector couplings on a particular bond acts.
		In this case, at any time only one bond is steered. For the operators acting on the same bond, $U_i^{(\ell,\ell+1)}U_j^{(\ell,\ell+1)\dagger}=0$ as the different $S^\tot=2$ states on each bond are orthogonal to each other.
		Hence, the inequality is satisfied in each of the steps. The evolution of the diagonal element of the system's density matrix corresponding to $\ket{\Psi_\tar}=\ket{\Psi_\aklt}$ is then described by
		\begin{widetext}
		\begin{equation}
		\braket{\Psi_\tar\vert\rho_\s(t+(\ell+1)\delta t)\vert\Psi_\tar}=\braket{\Psi_\tar\vert\rho_\s(t+\ell \delta t)\vert\Psi_\tar} + \underbrace{\sum_{i=1}^5 \braket{\Psi_\tar\vert U_i^{(\ell,\ell+1)}\rho_\s(t+\ell\delta t)U_i^{(\ell,\ell+1)\dagger}\vert\Psi_\tar}}_{Q_\ell}\sin^2\delta t,
		\label{eq:two-step-aklt}
		\end{equation}
		\end{widetext}
		where the equation corresponds to the specific case of steering the bond between sites $\ell$ and $\ell+1$.	The sum on the right-hand side of Eq.~\eqref{eq:two-step-aklt} is over all the detector qubits acting only on the bond between sites $\ell$ and $\ell+1$. Since $U_i^{(\ell,\ell+1)} U_j^{(\ell,\ell+1)\dagger} \neq0$, Eq.~\eqref{eq:two-step-aklt} follows from the derivation of Eq.~\eqref{eq:inequality-multiple-detectors} (Appendix~\ref{sec:derivation-multiple}). Note that $Q_\ell \geq 0$ for all $\ell$ as for $Q$ in Eq.~\eqref{eq:diagelem-time-evolved-exact}.
		In the time-continuum limit $\delta t\to 0$, the set of equations in \eqref{eq:two-step-aklt} for all $\ell$ give
		\begin{equation}
		\braket{\Psi_\tar\vert\rho_\s(t+N\delta t)\vert\Psi_\tar} =\braket{\Psi_\tar\vert\rho_\s(t)\vert\Psi_\tar} +  (\delta t)^2Q,
		\label{eq:aklt-timecont}
		\end{equation}
		with $Q=\sum_\ell Q_\ell$.

		Since $Q\ge0$, Eq.~\eqref{eq:aklt-timecont} ensures that the state is never steered further away from the target AKLT state if $Q$ is employed as a measure of distance from the target state. At the same time $Q=0$ does not necessarily ensure that the system is in the target state. As an illustration, consider an initial pure state that has excitations on multiple bonds. A single steering step of the type we have described removes one excitation without generating overlap with the target state, and so $Q=0$ for this step. So while the state changes, its overlap with the AKLT state does not.

		In fact, for our steering protocol it can be shown explicitly that there can arise $\rho_\s(t)$ for which the change in $E_\aklt(t)$ (defined in Eq.~\eqref{eq:EAKLT}) is positive. Using $E_\aklt(t)$ as a measure of distance one would then conclude that the state is steered further away from the AKLT ground state. To show this we use the fact that steering a bond affects only three bonds or equivalently the four sites involving them. Since $H_\aklt$ is a sum of projectors for each bond, the change in energy can be expressed to leading order in $\delta t$ as (see Appendix~\ref{sec:aklt-monotonic} for details)
		\begin{equation}
		\Delta E_\aklt(t) = -(\delta t)^2\mathrm{Tr}\left[\hat{\mathcal{E}}_\ell\rho_\s(t)\right],
		\end{equation}
		where $\hat{\mathcal{E}}_\ell$ is a four-site operator involving sites $\ell-1,\cdots,\ell+2$. Since the spectrum of the operator $\hat{\mathcal{E}}_\ell$ includes eigenvalues of both signs, monotonic decay of $E_\aklt(t)$ is not guaranteed. On the other hand, we find that for a choice of $\rho_\s(t)$ which leads to an initial growth of $\hat{\mathcal{E}}_\ell$, the distance from the target measured via $D_F(t)$ or $D_1(t)$ nevertheless decays monotonically. A detailed study of the monotonicity of steering, or lack thereof, depending on the distance measure, is presented in Appendix~\ref{sec:aklt-monotonic}. The upshot of this study is that one needs to show that the target state is the unique steady state of the dynamics. 
		As any state of the spin-1 chain which is not the AKLT state will have some overlap on the $S^\tot=2$ sector on one or more bond, the form of the operators in Eq.~\eqref{eq:spin1ops-steer-onto-stot=1} ensures that such a state is not a stationary state.
		
		To demonstrate that the AKLT state is indeed the unique state and also to show the validity of results away from the time-continuum limit, we simulate numerically the dynamics of a spin-1 chain starting from a random mixed state subjected to the steering protocol with all the bonds steered simultaneously. We calculate two quantities: (i) the reduced density matrix of a pair of adjacent spins-1
		which we denote as $\rho_\s^{(2)}$, and (ii) the distance of $\rho_\s(t)$ from $\rho_\aklt$ as a function of time, measured via $D_1$ and $D_F$.

		\begin{figure}
			\includegraphics[width=\columnwidth]{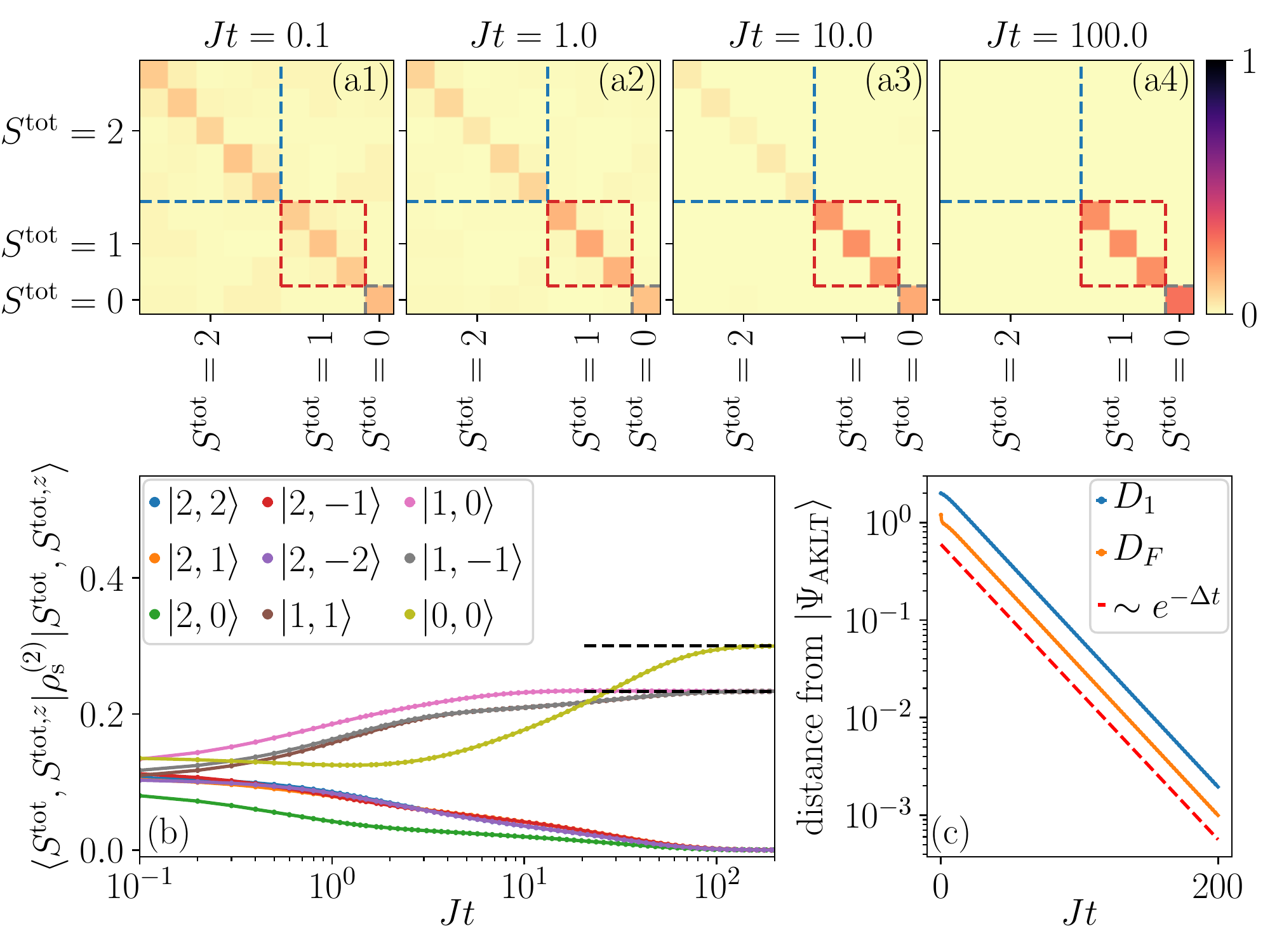}
			\caption{{\bf Steering of a spin-1 chain to the AKLT ground state.} (a) The evolution of the reduced density matrix of two adjacent spins, $\rho_\s^{(2)}$, shown as a colourmap (for the absolute values of the matrix elements) in the $\ket{S^\tot,S^{\tot,z}}$ basis. The support on $S^{\tot}=2$ sector decays to zero and so do the off-diagonal elements in this basis. (b) The evolution of the diagonal elements of $\rho_\s^{(2)}$. The horizontal dashed lines denote the values 1/3 and 2/9, obtained from $\rho_\aklt^{(2)}$. (c) The distance between $\rho_\s(t)$ and $\rho_\aklt$ as measured via the matrix norms $D_F$ and $D_1$ decays exponentially to zero with time. The dashed line represents an exponential decay with a rate $\Delta$ obtained from the spectral gap of the corresponding Lindbladian. For the plots, $N=5$, $J=1$, and $\delta t=0.1$. }
			\label{fig:aklt-discrete-time}
		\end{figure}

		The results are shown in Fig.~\ref{fig:aklt-discrete-time}.
		The evolution of $\rho_\s^{(2)}(t)$ shows that each bond is indeed steered out of the $S^\tot = 2$ subspace.
		However, a drastic difference between the dynamics of $\rho_\s^{(2)}$ and that of a single pair of spins-1 is that the former has a unique steady state which is diagonal in the $\ket{S^\tot,S^{\tot,z}}$ basis with $\braket{1,s_1|\rho_\s^{(2)}(t\to\infty)|1,s_1}=2/9$ and $\braket{0,0|\rho_\s^{(2)}(t\to\infty)|0,0}=1/3$.
		This highlights the collective effect of the many-body yet local steering protocol, as each of the system-detector couplings are such that they keep the off-diagonal elements within the $S^\tot=1$ and $0$ subspace invariant.
		Moreover, note from the terms in Eq.~\eqref{eq:spin1ops-steer-onto-stot=1} that they do not change the diagonal element corresponding $S^\tot=0$.
		On the other hand, in the many-body case, the matrix element of $\rho_\s^{(2)}$ on a particular bond goes to its steady state value of $1/3$ purely due to the interplay of the steering on the bond with that on the other bonds, further highlighting the collective effects of the many-body steering protocol.

		The fact that $\rho_\s^{(2)}$ approaches a diagonal form in the $\ket{S^\tot,S^{\tot,z}}$ basis with the diagonal elements taking the above mentioned specific values is crucial because $\rho_\aklt^{(2)}$ has exactly this feature.
		For the reduced density matrix of a subsystem of the spin-1 chain comprising the contiguous sites $\ell_1$ through $\ell_n$, it can be easily argued that its eigenvectors corresponding to non-zero eigenvalues are the four degenerate AKLT ground states of the spin-1 chain of length $n$ with open boundary conditions described the Hamiltonian $H_\mathrm{subsys}=\sum_{\ell=\ell_1}^{\ell_n}  \mathcal{P}_{S^\tot=2}^{(\ell,\ell+1)}$. 
		At the same time, $H_\mathrm{subsys}$ commutes with $(\sum_{\ell=\ell_1}^{\ell_n} \hat{\mathbf{S}}_\ell)^2$ and $\sum_{\ell=\ell_1}^{\ell_n} \hat{S}^z_\ell$.
		These eigenstates are simultaneous eigenstates of $H_\mathrm{subsys}$, $(\sum_{\ell=\ell_1}^{\ell_n} \hat{\mathbf{S}}_\ell)^2$, and $\sum_{\ell=\ell_1}^{\ell_n} \hat{S}^z_\ell$.
		For $n=2$, it directly implies that $\rho^{(2)}_\s$ is diagonal in the $\ket{S^\tot,S^{\tot,z}}$ basis. Moreover, it has been shown that the four non-zero eigenvalues are $(1+3(-3)^{-n})/4$ and $(1-(-3)^{-n})/4$ with a threefold degeneracy~\cite{korepin2010entanglement} which for $n=2$ are 1/3 and 2/9 respectively.
		Since every bond of the spin-1 chain is steered to a state that corresponds to the AKLT ground state of the entire chain, it is natural that the many-body state $\rho_\s(t)$ is also steered to $\rho_\aklt$. 
		This confirmed by the exponential decay of the distances, $D_F$ and $D_1$ between $\rho_\s(t)$ and $\rho_\aklt$ with time as shown in Fig.~\ref{fig:aklt-discrete-time}(c).

		\begin{figure}
			\includegraphics[width=\columnwidth]{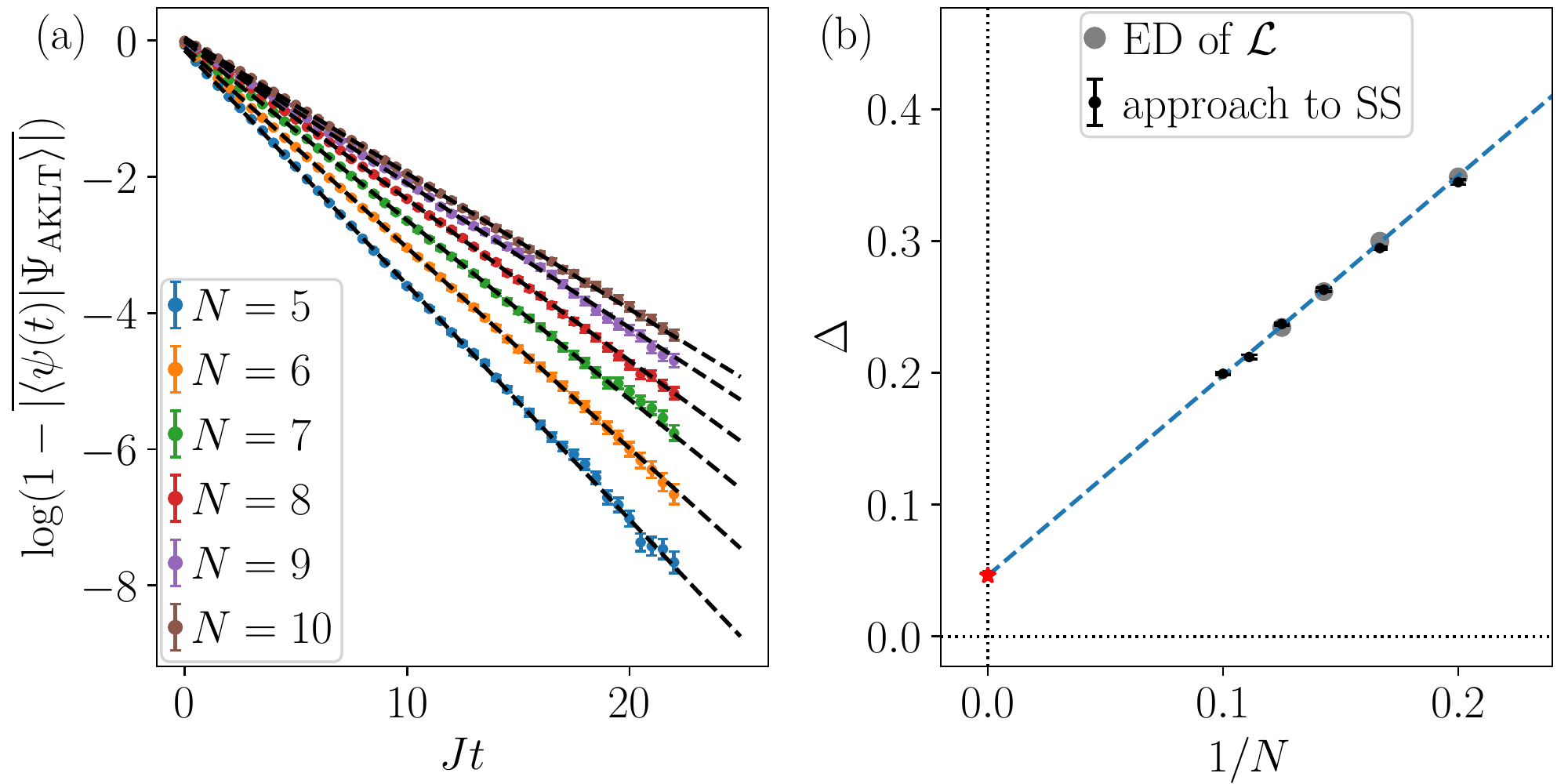}
			\caption{{\bf Scaling of steering rate with system size for the spin-1 chain.} (a) The exponential approach of the spin-1 chain to the AKLT state as obtained from the wave-function Monte Carlo method for solving the Lindblad dynamics for different system sizes $N$. As the method involves evolving pure states, we measure the distance from the AKLT state via $\log(1-\overline{\braket{\psi(t)|\Psi_\aklt}})$ where the bar denotes the average over the wave-function trajectories. The black dashed lines are fits to the data used to extract $\Delta$. (b) The rate of the exponential approach to the steady state (SS) obtained from (a) or equivalently the gap in the spectrum of the Lindbladian obtained from exact diagonalisation of $\mathcal{L}$ shown as a function of $1/N$. The intercept indicates a finite $\Delta$ in the thermodynamic limit. The data for panel (a) was averaged over 20000 trajectories and the errorbars were obtained via standard bootstrap with 500 resamplings. Errors in the extrapolated value of $\Delta$ introduced via the errors in $\Delta$ for different $N$ as well as the fitting procedure are also shown but the error bars are smaller than the point size.}
			\label{fig:aklt_mc_stot=1}
		\end{figure}

		So far we have argued that $\rho_\aklt$ is a steady state of the steering protocol and have shown numerically that an arbitrary initial state of a finite spin-1 chain evolved with the steering protocol approaches the AKLT ground state exponentially fast in time.
		The natural question to ask then is what happens to the steering protocol in the $N\to\infty$ limit, with regard to both the uniqueness of the steady state and the time-scale for steering to it.
		We find it most convenient to answer these questions in terms of the Lindbladian corresponding to the steering protocol.
		Following the discussion in Sec.~\ref{sec:lindblad}, the jump operators, of which there $5\times N$, can be immediately identified as $L_i^{(\ell,\ell+1)} = J U_i^{(\ell,\ell+1)}\sqrt{\delta t}$.
		For convenience, we will set $J \sqrt{\delta t}=1$ in the following.
		As discussed in Sec.~\ref{sec:lindblad}, the uniqueness of the steady state manifests itself in the non-degeneracy of the zero eigenvalue of $\dvec{\mathcal{L}}$. The rate of the approach to the steady state is given by the gap $\Delta$ of the rest of the spectrum from the zero eigenvalue.
		Hence, it is instructive to study the scaling of $\Delta$ with $N$.
		
		By exactly diagonalising $\dvec{\mathcal{L}}$ for finite sizes we find that the zero energy eigenvalue is indeed non-degenerate and the unique steady state corresponding to it is identical to $\rho_\aklt$.
		The dimension of $\dvec{\mathcal{L}}$ is $3^{2N}$, which restricts calculations to $N\le 8$.
		In order to extract $\Delta$ for slightly larger sizes, $N\le 10$, we solve for the dynamics of the Lindblad equation using the \emph{wave-function Monte Carlo} method which involves evolving pure states with stochastic quantum jumps and then averaging over the so-obtained trajectories~\cite{dalibard1992wave,molmer1993monte}.
		With this method, we find again that the spin-1 chain approaches the AKLT ground state exponentially fast in time and hence the rate $\Delta$ can be extracted, see Fig.~\ref{fig:aklt_mc_stot=1}(a).
		The $\Delta$ obtained from the dynamics for smaller system sizes agrees very well with the one obtained from diagonalising $\dvec{\mathcal{L}}$, see Fig.~\ref{fig:aklt_mc_stot=1}(b).
		The two methods together allow us to study the scaling of $\Delta$ with $N$ and from the scaling shown in Fig.~\ref{fig:aklt_mc_stot=1}(b), we conclude
		\begin{equation}
			\Delta(N) = \Delta_\infty + c/N\,.
			\label{eq:Deltascaling}
		\end{equation}
		Hence the gap in the the spectrum of $\dvec{\mathcal{L}}$ or equivalently the time-scale required to steer the spin-1 chain to the AKLT state stays finite in the thermodynamic limit.

		Combining all the results presented in this section, we conclude that the map of Eq.~\eqref{eq:map}, with the detectors' initial states and the system-detector coupling described by Eqs.~\eqref{eq:detector-intialise-aklt} and \eqref{eq:hamcoupling-aklt} respectively, constitutes a protocol for steering a many-body system to a strongly correlated state, in this case a spin-1 chain to the AKLT ground state. The protocol satisfies the constraints of having an extensive number of detector qubits coupled only locally to the system and of having a steering time-scale that does not diverge in the thermodynamic limit.

\section{Discussion \label{sec:discussion}}

In this work we have developed a general protocol that uses entanglement to induce steering of a quantum system to a target state. We have shown how our protocol can be applied both to steer systems with a few degrees of freedom and to prepare strongly correlated states in many-body systems. The form of steering that we study here is rooted in but distinct from steering of the type discussed originally in the context of the EPR paradox~\cite{schrodinger1935discussion,schrodinger1936probability}. Specifically, our protocol shares with earlier discussions of quantum steering the fact that it exploits the entanglement between the system of interest and another system. Crucially, however, it differs in offering a general framework to steer a quantum system into a specific and pre-designated state.

\subsection{Experimental realisations}
Experimental implementations of the ideas presented in this work require realisations of 
qubits which satisfy several conditions. (i) The relaxation time of the qubit due to undesired coupling to external degrees of freedom should be long. (ii) Likewise, the coherence times 
should be long. (iii) It should be
possible to design system-detector couplings as required for versatile steering to arbitrary target states. (iv) The system should be scalable to multiple qubits and should allow for multi-qubit interactions (for example a three qubit-interaction term in the system-detector Hamiltonian involving one detector qubit and two system qubits).

There are several promising candidate platforms for realising our protocol, which include superconducting qubits, cold ions, nitrogen-vacancy (NV) centres, macroscopic mechanical objects with entangled quantum degrees of freedom, and quantum optics systems.
These are platforms that have all been tested, or will be tested, as far as the steering of a single qubit is concerned. The extent to which more complex systems, comprising multiple quantum degrees-of-freedom, may be steered relies on the extent to which conditions listed earlier can be satisfied. In the following we briefly discuss some of these platforms.

\paragraph*{Superconducting qubits:} The most prospective experimental avenue for our proposal is based on superconducting qubits. There are two realizations of qubits which   appear to be the leading candidates for constructing  steering platforms. The first are {\it transmons}. In previous experimental work, weak measurements were constructed via entangling operations between two superconducting qubits, and subsequent projective measurement on the detector qubit~\cite{groen13} which is exactly the measurement structure we propose in this work.  Another approach, implemented primarily for the steering of a single qubit, relies on dispersive interaction between the qubit and a coupling to a waveguide cavity. Existing implementations of weak measurement back-action on such a systems~\cite{hatr13,murch2013observing, tan15, HacohenGourgy2016} have three major differences from our steering protocol. First, the sequence of weak measurements in the experiments amounts to a strong measurement when integrated over time (i.e., when considering a sequence of many weak measurements), which projects the initial state onto a final state with probabilistic outcomes. Second, the experiments make use of post-selected readouts. Third, the specifics of the experiment rely on knowledge of the system’s initial state. A further experiment~\cite{nagh19,minev2019catch} focusses on the study of a three level system, where the evolution of a two-level subsystem is conditioned on the system not being in the third level. This can be represented using modified Lindblad dynamics leading to non-hermitian Hamiltonian dynamics. Alternatively, the dynamics can be described in terms of a set of null weak measurements with post-selected readout sequence~\cite{zilberberg2012null,zilberberg2013null}. Either picture differs substantially from our protocol. However, adjusting these protocols to our paradigm of steering should be rather straightforward.  Generalizations of these protocols could work for a detector coupled to two or more qubits~\cite{murch-pc}.  

The second possible realization of superconducting qubits involves the use of state-of-the-art fluxonium qubits. These offer coherence times comparable to more common transmon qubits, with increased anharmonicity, reduced cross-talk, and design flexibility. Such devices could be implemented to construct both our many-body system as well as the detector qubits~\cite{pop-pc,manucharyan2009fluxonium,pop2014coherent,grunhaupt2019granular}. Concretely, two fluxonium qubits could be coupled with variable strengths. One fluxonium plays the role of the qubit to be steered, while the other fluxonium, inductively coupled to a readout resonator~\cite{pop2014coherent}, plays the role of detector on which dispersive projective readout can be performed with high fidelity. The coupling between the qubits can be implemented using shared capacitors or shared inductors. With the recent availability of compact inductors with low nonlinearity, using disordered superconductors such as granular aluminum~\cite{maleeva2018circuit} or TiN~\cite{shearrow2018atomic}, inductive coupling appears more appealing, as it minimizes cross-talk between the qubit to be steered and the readout resonator.

\paragraph*{Cold ions:} Another prospective platform for implementing measurement-induced steering is based on cold ions trapped by magnetic field and lasers. Qubits are implemented in 
electronic states of each ion and the ions in a common trap can communicate with each other
via intrinsic long-range interactions and external, laser-induced couplings. Below, we will describe an example of such a platform consisting of Sr\textsuperscript{+} ions which are essentially hydrogen-like atoms with one outer
electron; the lowest two levels represent the qubit. 
Coupling to the ion with light can effect arbitrary unitary rotations. 
Qubit operations are well-developed in this area, but steering of a single qubit is only underway, while a challenge for applications to steering is to develop multiple-qubit operations.

Generalizing from single-qubit to two-qubit measurements is highly challenging. The more straightforward protocol would involve shining light of a wavelength larger than the distance between ions, so that each ion can absorb and emit a photon, ensuring coherence between the two ions~\cite{akerman2012reversal,glickman2013emergence,lin2013dissipative,kienzler2015quantum}.
The 
detection of 
an  emitted  photon in a given direction  determines the backaction.  To achieve versatility of the system-detector Hamiltonian, one needs to go one step further and designate one ion as a detector that measures two other ions (one on the left and the other on the right of it). Every two neighbouring ions can be made to interact through terms of the system-detector Hamiltonian that flip the pair of qubits. If it is possible to couple and decouple on-demand  the detector and the left spin, and separately the detector and the right spin, then performing unitary transformations on the right spin-detector and/or the left spin-detector, a broad class of effective 3-qubit Hamiltonians can be achieved.

\paragraph*{Nitrogen-vacancy (NV) centres:} NV centres in diamonds are point defects which realise electronic spins localised at atomic scales. In these setups, the detector could be the NV centre and the qubits could be {}\textsuperscript{13}C nuclei spins.
The interaction between them is of a dipole-dipole type, which may be controlled with high precision~\cite{retzker-pc}. Direct optical  measurement can only be preformed on the NV center.  The control is made with microwaves~\cite{cai2013diamond}.
Another possibility is to use NV centres to control external spins~\cite{cai2013large}. In principle, one could aim at a multi-spin measurement or control. The level structure for such systems has been analysed extensively~\cite{maze2011properties}. Various aspects of the NV readout and polarization mechanism~\cite{batalov2008temporal}
as well as control at weak coupling have been studied (cf. Refs.~\cite{gefen2018quantum} and \cite{cujia2019tracking} for theory and experiment respectively). We note that the control can be much improved at low temperatures~\cite{abobeih2019atomic}.

\paragraph*{Macroscopic mechanical objects:} Measuring and controlling macroscopic mechanical objects that exhibit quantum coherence is a challenge realised in recent years~\cite{rossi2018measurement}. The next step, accomplished very recently~\cite{kotler2020tomography}, is to scale this up to two macroscopic mechanical degrees of freedom. Unfortunately, achieving a high level of control and measurement may expose the system to undesired interactions with its environment, a problem that becomes more pronounced at the macroscopic scale. Using a superconducting electromechanical circuit and a pulsed microwave protocol, a system of two mechanical drumheads  
has been steered towards an entangled state, where entanglement of continuous degrees of freedom satisfies the criteria formulated~\cite{simon2000peres,duan2000inseparability}. The strength of the measurement %here
in these systems
can be tuned from strong to weak, and  coherence may be maintained over long time-scales. 
However, given the immense experimental difficulties here, the issues of versatility of system-detector  Hamiltonian, and the question of scalability to systems consisting of multiple degrees of freedom 
are all
likely to remain a challenge in the near future.

Finally, one could adopt these ideas to the field of quantum optics~\cite{karimi-pc}. Photon polarization has been used as a qubit, for example, in the process of  storing  and then retrieving quantum information in cold atom platforms~\cite{wang2019efficient}.  More directly related to our ideas are two polarization-based  qubits (one serving as “system” and the other as “detector”) allowing for back-action control~\cite{cho2019emergence}. 

In summary, experimental implementations of our ideas should be assessed according to the complexity of the required platform. As far as a single-qubit steering is concerned, reliable implementations are achievable with all platforms. Entangling two qubits has already been done with some platforms, with the promise that other types of manipulations are, in principle, also possible. Going to the arena of many-body physics, one needs to engineer scalable multi-qubit states by our steering protocols. This is feasible with platforms based on superconducting qubits and NV centres, where flexible
(and, in principle, arbitrarily complex) few-qubit-detector couplings can be engineered, as demonstrated for two-qubit realizations. Implementation of many-body steering on these platforms requires experimental efforts towards long coherence times, circuit connectivity, and quality control of system-detector coupling and detector re-initaition. Importantly, 
even when some specific system-detector couplings are still hard to design for technological reasons, the shake-and-steer protocol may resolve this issue. We expect our work to stimulate experiments based on existing and future platforms in the direction of steering and manipulating many-body states.

\subsection{Connection to previous work}
The use of external degrees of freedom as a designed environment for manipulating and controlling quantum states has been discussed previously in other settings~\cite{zhang2017quantum,altafini2012modeling,ticozzi2012stabilizing,ticozzi2017alternating}. Treatment of the environment as Markovian or the measurements as non-selective leads to a Lindblad equation, similar to one derived in the present work. Much of the earlier focus has been either  on the formal aspects of such non-unitary dynamics, or on applications involving few-body quantum systems. Our work 
constitutes a generalisation of
such quantum control and steering to macroscopic many-body systems.

In the context of open quantum systems, the creation of non-trivially correlated or topological many-body states has been proposed using so called \emph{drive-and-dissipation} schemes~\cite{diehl2008quantum,kraus2008preparation,roncaglia2010pfaffian,diehl2011topology,bardyn2013topology,leghtas2013stabilizing,liu2016comparing,goldman2016topological}. 
In such schemes the system is driven or excited using a time-dependent Hamiltonian while a dissipative channel, often a Markovian environment, working simultaneously relaxes the system. This interplay of drive and dissipation is then used to engineer non-trivial states. From a theoretical point of view, the Markovian nature of the environment naturally lends itself to a treatment via the Lindblad equation, hence yielding a formal connection to our measurement protocol in the time-continuum limit alluded to previously. In fact, the jump operators \emph{postulated} in Ref.~\cite{kraus2008preparation} in the context of the AKLT chain also bear formal similarities with those derived in our case from the measurement protocol. 

Further related work is described in
Ref.~\cite{weimer2010rydberg}, where auxiliary qubits are exploited to drive a Rydberg-atom realisation of Kitaev's toric code to its ground state. Excited states of the toric code contain spatially localised excitations on plaquettes or vertices of the system, which can annihilate in pairs. 
The protocol of Ref.~\cite{weimer2010rydberg} takes advantage of this special feature of the toric code.
To cool this system to its ground state, an implementation of 
Lindblad jump operators is proposed that 
generates stochastic motion of excitations, so that pairs meet and annihilate.
An advantage of our steering protocol for the AKLT chain is that 
we directly steer each local degree of freedom of the system (a bond between two sites in the AKLT chain) towards the ground state. 
Moreover, our general protocol is independent of special features of the steered system. The local unit which is steered could have a large excited state manifold; steering would still be possible using the shake-and-steer protocol (see Appendix~\ref{sec:shake-and-steer}) with a finite number of locally coupled detectors. 

In contrast to the drive-and-dissipation protocols, our measurement-based protocol does not require a system Hamiltonian to induce dynamics in the system. Moreover, unlike an environment which is to a large extent uncontrolled, the detectors that play the role of the dissipative channel are well-controlled simple quantum systems that can be tuned in time. As on-demand jump operators, they 
represent a particularly 
a useful form of quantum manipulation and control. Moreover, as outlined above, using the measurement outcomes for active decision making, or having a measurement protocol which varies with time and has a memory kernel, takes us to realms that seemingly cannot be addressed by drive-and-dissipation techniques.

\subsection{Future directions}
There is a wide range of possible future directions and open questions arising from the work presented here, and we close by outlining some of these.
While the protocol we have described is arguably the simplest version, it is also quite robust in the sense that the probability of reaching the target state is unity irrespective of the initial state of the system. One can also envisage a wide variety of generalisations, in which instead of simply decoupling the detector qubits from the system of interest at the end of each steering step, a projective measurement is made of the final detector states. In this broader context, the protocol we have described corresponds to the special case of blind measurements in which an unbiased average is made over all such measurement outcomes. A natural direction is to ask whether making use of the measurement outcomes can allow optimisation of the protocol. More precisely, the initial state of the detector, the system-detector coupling, and the duration of the coupling at each step can be made to depend on the measurement outcome of the previous step or on an extended history of the measurement outcomes. Active error correction could be viewed as an example of a broader class of protocols involving such decision making. Generalised protocols of this type are manifestly not Markovian and hence are not describable in terms of a Lindblad equation with time-independent jump operators. In the context of quantum control~\cite{zhang2017quantum}, there have been proposals to utilise the measurement outcomes as a feedback to the dynamics using the so-called \emph{quantum filtering equations}~\cite{belavkin1992quantum,altafini2012modeling}. Another direction for taking non-Markovian effects into account is via the quantum collision models~\cite{ciccarello2017collision}. We reserve the generalisation of our protocol to include such effects for future work.

There are many possible motivations for such generalisations. One would be to optimise the time-scale over which the system reaches a pre-defined vicinity of the target state. Another would be to maximise the purity of the quantum state of the system throughout the protocol. This would become particularly important if one extends the objective of the steering protocol from preparation of a target state to manipulation of a state using a sequence of weak measurements~\cite{gebhart2019topological}.  

Another set of open questions concerns the robustness of the protocol to deformations of the initial states of the detector qubits or of the system-detector couplings. These questions call for a detailed and comprehensive study of the consequences of errors in steering protocols and possible ways of correcting them, which we leave for a future work. Specifically, one could ask under these conditions whether the system has a steady state, and if so how its distance from the target state scales with the strength and probability of errors in the steering protocol. Furthermore, one might look for specific error correction protocols akin to a stabiliser formalism~\cite{gottesman1997stabilizer}, but within the setting of a steering protocol.

It is reasonable to anticipate that a finite gap in the spectrum of the Linbladian in the thermodynamic limit (as we find in our study of the AKLT state) will ensure that the steady state is robust to small random errors in the steering protocol that are local in space and time. Robustness of the protocol to time-independent deformations is a separate issue. A specific instance concerns the interplay of a system Hamiltonian (which we have so far omitted) with the system-detector coupling Hamiltonian. This would amount to having a non-zero $V$ in Eq.~\eqref{eq:lindblad-derived}. It is clear that if the Lindbladian in the absence of the system Hamiltonian shares an invariant subspace with the system Hamiltonian,
the steering protocol is robust to the latter.
However, if that is not the case, the apparent tension between the system Hamiltonian and the system-detector coupling Hamiltonians could lead to a new steady state which deviates continuously with the strength of the system Hamiltonian from the original dark state. Alternatively, in the many-body case there could be a dynamical phase transition in the nature of the steady state.

In fact, measurements on an otherwise random unitary circuit have been shown to induce an entanglement phase transition in the steady state as a function of the density or strengths of measurements~\cite{li2018zeno,skinner2019measurement,chan2019unitary,li2019measurement,szyniszewski2019entanglement,jian2019measurement,gullans2019dynamical,choi2019quantum,bao2019theory,tang2019measurement}. Most of these works use local projective measurements or generic positive operator-valued measurements which manifestly decrease the entanglement. However, our work shows that measurement protocols can be constructive in the sense that they can be designed so that steady state is a strongly correlated state. Thus a natural question arises as to the extent  to which measurement protocols and their interplay with system Hamiltonians can be classified in terms of entanglement properties of the resulting steady states.

An intriguing future direction would be to study measurement protocols for which the emergent Lindbladian has a \emph{dark space} spanned by several dark states. An example is provided by the AKLT state in an open chain, for which this space is four dimensional. In such a scenario one can envision a closed adiabatic trajectory (in the protocol parameter space) that could be used to induce a unitary transformation in the dark space. This could be harnessed to induce adiabatic rotations in a degenerate many-body space, and may even give rise to a many-body non-Abelian geometric phase~\cite{wilczeck1984appearance,snizhko2019nonabelian}. Similarly, the presence of multiple stationary states of Lindbladians has been used theoretically to realise dissipative time-crystals~\cite{gong2018discrete,gambetta2019discrete,lazarides2019time,osullivan2018dissipative}, and these could in principle also be generated via a measurement protocol.

\begin{acknowledgments}
E. Karimi, K. Murch, R. Ozeri, I. Pop, A. Retzker, and K. Snizhko have made insightful comments on experimental realisations. We have benefited from useful discussions with Y. Herasymenko, P. Kumar, and K. Snizhko. We acknowledge financial support from EPSRC Grant Nos. EP/N01930X/1 and EP/S020527/1, from the Israel Science Foundation, from the Leverhulme Trust Grant VP1-2015-005, and from the Deutsche Forschungsgemeinschaft (DFG) Projektnummer 277101999, TRR 183 (project C01), and Grants No. EG 96/13-1 and GO 1405/6-1.
\end{acknowledgments}

\appendix

\section{Steering a single spin-1/2 to an arbitrary state \label{sec:arbitrary}}

\begin{figure}
\includegraphics[width=0.75\columnwidth]{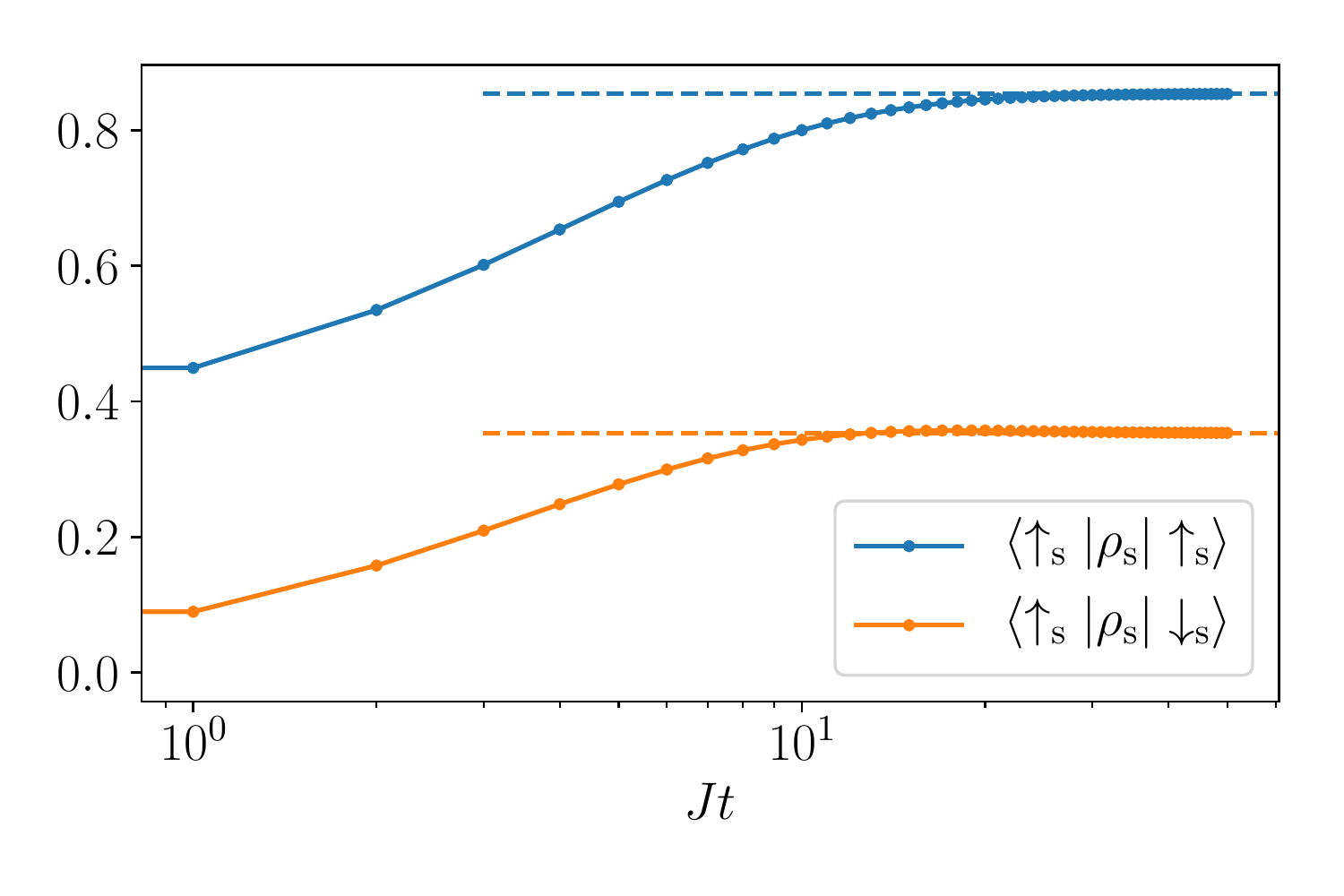}
\caption{Steering of a single spin-1/2 using the system detector coupling in Eq.~\eqref{eq:hsddiff} with $\theta=\pi/4$. The different lines show the different matrix elements of the system spin's density matrix. The horizontal dashed lines show the values that the matrix elements should take for $\rho_\s=\ket{\Psi_\tar}\bra{\Psi_\tar}$.}
\label{fig:diffdir}
\end{figure}

In the toy example of a single spin-1/2 presented in Sec.~\ref{subsec:toy}, the target state of the system qubit was the same as the initial state of the detector. In this appendix, we demonstrate that this is not a requirement, and show that our protocol can be used to steer the system qubit to an arbitrary target state. Let us suppose that we always prepare the detector spin in its $\ket{\uparrow_\a}$ state, but we want to steer the system spin to a pure state of the form
\begin{equation}
\ket{\Psi_\tar} = \cos(\theta/2)\ket{\uparrow_\s}+\sin(\theta/2)\ket{\downarrow_\s}.
\end{equation}
This is easily achieved using a system detector coupling Hamiltonian of the form
\begin{equation}
\hsd =  J\sigma^-_\a \otimes \ket{\Psi_\tar}\bra{\Psi_\tar}(-i\sigma^y_\s)+ \mathrm{h.c.},
\end{equation}
where $-i\sigma^y_\s\ket{\Psi_\tar}$ is the state orthogonal to $\ket{\Psi_\tar}$ and out of which we want to steer. In terms of the spin operators, the Hamiltonian can be written as
\begin{equation}
\hsd=\frac{J}{2}(\sin\theta\,\sigma^z_\s-\cos\theta\,\sigma^x_\s-i\sigma^y_\s)\otimes\sigma^-_d + \mathrm{h.c.}
\label{eq:hsddiff}
\end{equation}
The resulting dynamics of $\rho_\s$ under the protocol with the system-detector coupling Hamiltonian as above is shown in Fig.~\ref{fig:diffdir} for $\theta = \pi/4$. The different lines correspond to the matrix elements of the time-evolving density matrix of the spin with $\braket{\uparrow|\rho_\s(t\to\infty)|\uparrow} = \cos^2(\theta/2)$ and $\braket{\uparrow|\rho_\s(t\to\infty)|\downarrow} = \cos(\theta/2)\sin(\theta/2)$ at long times showing that in the steady state $\rho_\s\to\ket{\Psi_\tar}\bra{\Psi_\tar}$.

\section{Derivation of steering inequality with multiple detector qubits \label{sec:derivation-multiple}}

In this appendix, we present some detail of the derivation of the steering inequality in the presence of multiple detectors. Partitioning the composite Hilbert space of the system and the detectors into $\rho_\a\otimes \mathcal{H}_\s$ and $(\overline{\mathcal{D}}_\a)\otimes\mathcal{H}_\s$ as in Sec.~\ref{sec:map}, the system-detector coupling Hamiltonian of Eq.~\eqref{eq:hamcoupling} can be expressed as 
\begin{widetext}
\begin{equation}
\hsd = \bordermatrix{&\ket{\Phi_\a} &\cdots& O_\a^{(1)}\ket{\Phi_\a} & \cdots  & O_\a^{(n)}\ket{\Phi_\a} &\cdots\cr
                \ket{\Phi_\a}& 0 &  0& U_1^\dagger  & \cdots & U_n^\dagger & \cdots &\cr
                \vdots& 0 &  0& \cdots  & \cdots & \cdots & 0 &\cr
                O^{(1)}_\a\ket{\Phi_\a}& U_1  &  \vdots & \ddots & \ddots &\ddots & \vdots\cr
                \vdots& \vdots  &  \vdots & \ddots & \ddots &\ddots & \vdots\cr
                O^{(n)}_\a\ket{\Phi_\a}& U_n  &  \vdots & \ddots & \ddots &\ddots & \vdots\cr
                \vdots& \vdots  &  0 & \ddots & \ddots &\ddots & 0\cr},
\label{eq:W-multiple}
\end{equation}
where the labels $\{O_\a^{(n)}\ket{\Phi_\a}\}$ on the rows and columns of the matrix denote the positions of non-zero entries, $U_n$ and $U_n^\dagger$. The entry labelled by $n$ corresponds to the coupling of the system with the $n^\mathrm{th}$ detector.
With this notation, the operator $W=\exp\left(-i\hsd\delta t\right)$ takes the form
\begin{equation}
W = \sum_{k=0}^\infty\frac{(-i\delta t)^{2k}}{(2k)!}
\begin{pmatrix}
X^k & \frac{-i\delta t}{2k+1}X^k[\cdots ~ U_1^\dagger ~ \cdots ~ U_n^\dagger ~ \cdots]\\
\frac{-i\delta t}{2k+1}Y^k\begin{bmatrix}
0 \\
U_1\\
\vdots\\
U_n\\
\vdots
\end{bmatrix}& Y^k
\end{pmatrix},
\end{equation}
where
\begin{equation}
X = \sum_{i}U_n^\dagger U_n; ~~~ Y = \begin{bmatrix}
\vdots \\
U_1\\
\vdots\\
U_n\\
\vdots
\end{bmatrix}\otimes[\cdots ~ U_1^\dagger ~ \cdots ~ U_n^\dagger ~ \cdots].
\end{equation}
Using the form of the operator $W$ from Eq.~\eqref{eq:W-multiple} in the map described by Eqs.~\eqref{eq:evol} and \eqref{eq:map}, we obtain the time-evolved density matrix of the system as
\begin{equation}
\rho_\s(t+\delta t) = \sum_{k,l=0}^\infty \frac{(-i\delta t)^{2k}(i\delta t)^{2l}}{(2k)!(2l)!}X^k\rho_\s(t)X^l
+
\sum_{k,l=0}^\infty \frac{(-i\delta t)^{2k+1}(i\delta t)^{2l+1}}{(2k+1)!(2l+1)!}\sum_n U_nX^k\rho_\s(t)X^l U_n^\dagger.
\label{eq:multi-rho-evol}
\end{equation}
The quantity of interest is $\braket{\Psi_\tar\vert\rho_\s(t)\vert\Psi_\tar}$, which can be obtained from Eq.~\eqref{eq:multi-rho-evol} under the condition that either $U_\s^{(m)}U_\s^{(n)\dagger} = 0$ for $n\neq m$ or $U_\s^{(m)}$ and $U_\s^{(n)\dagger}$ commute, in the form
\begin{equation}
\braket{\Psi_\tar\vert\rho_\s(t+\delta t)\vert\Psi_\tar}=\braket{\Psi_\tar\vert\rho_\s(t)\vert\Psi_\tar} + \sum_n \braket{\Psi_\tar \vert U_\s^{(n)}\rho_\s(t) U_\s^{(n)\dagger}\vert\Psi_\tar}\sin^2(\delta t),
\label{eq:inequality-multiple-detectors}
\end{equation}
\end{widetext}
which is the same as equation as Eq.~\eqref{eq:diagelem-time-evolved-exact} with $Q$ given by Eq.~\eqref{eq:Q-multiple}. 
This concludes our derivation of the steering inequality for the case of multiple detectors.

\section{Shake-and-steer protocol \label{sec:shake-and-steer}}

\begin{figure}
	\includegraphics[width=\columnwidth]{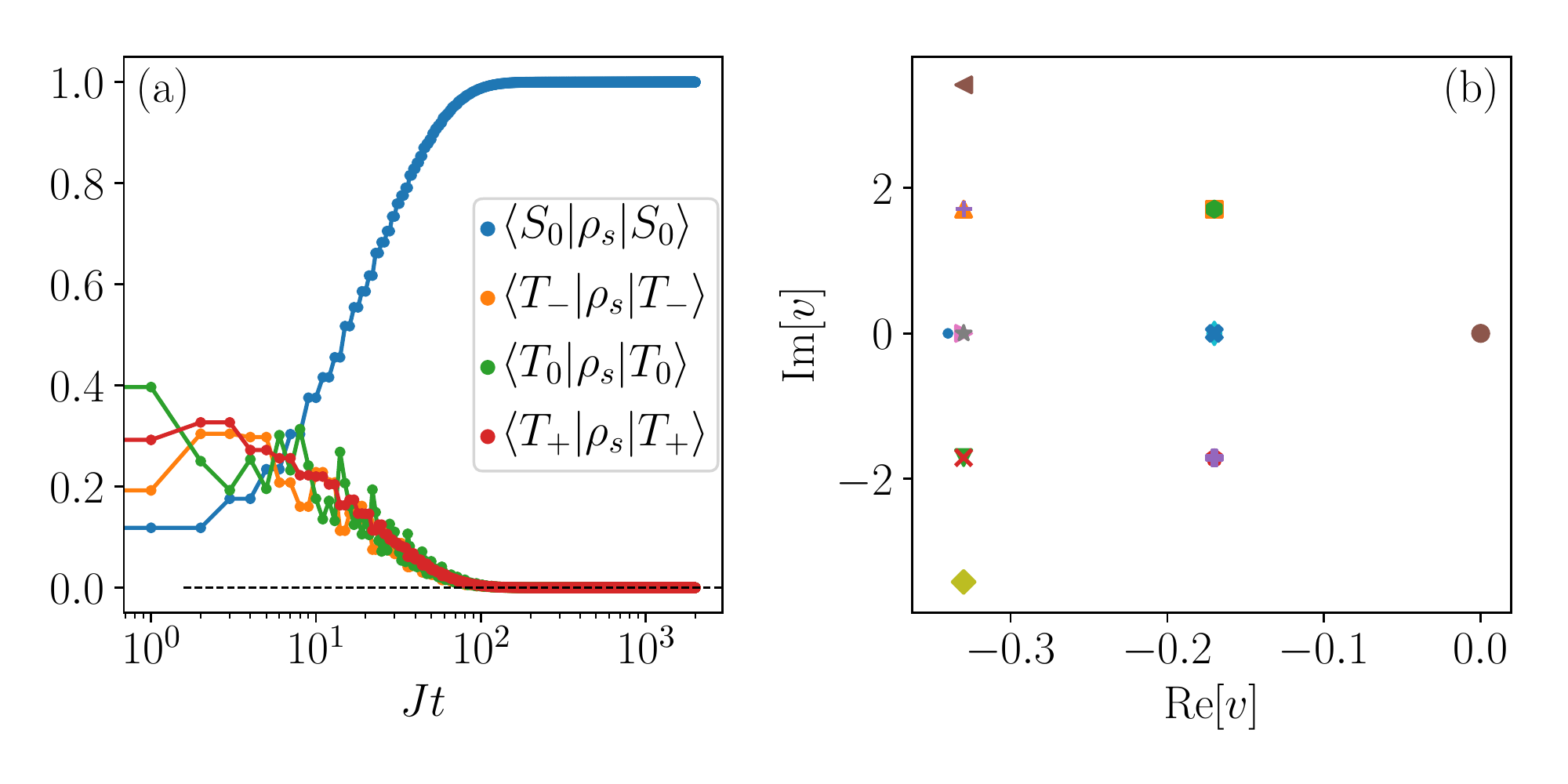}
	\caption{\textbf{Shake-and-steer protocol for steering a pair of spins-1/2 to the singlet state.} (a) The dynamics of the diagonal elements of the density matrix under the protocol described in Eq.~\eqref{eq:shake-and-steer-singlet-protocol} with $\bm{\theta_{1}}=\bm{\theta_{2}}=J(1,1,1)$. (b) The spectrum of the corresponding Lindbladian with a unique zero-eigenvalue corresponding to the singlet state.}
	\label{fig:shake-and-steer-singlet}
\end{figure}

In all the previously discussed cases, the number of system-detector couplings used locally is equal to the dimension of the subspace orthogonal to the target state. The shake-and-steer protocol permits steering with just one system-detector coupling. An additional Hamiltonian acting on the system unitarily rotates the state of the system within the orthogonal subspace. Thus the weight from the orthogonal subspace keeps leaking into the target state via one channel corresponding to the system-detector coupling, while the weight from the rest of the orthogonal subspace is \emph{shaken} into the dissipative channel via the system Hamiltonian. Let us illustrate the protocol using the steering of two spins-1/2 to the singlet state, similar to the one in Sec.~\ref{sec:spin1/2}.

Out of the three system-detector couplings used there, Eq.~\eqref{eq:hamcoupling-singlet-terms}, let us consider only one of them, such that
\begin{equation}
	\hsd = J(\sigma^-_\a \otimes U_3 + \mathrm{h.c.}).
	\label{eq:shake-and-steer-spin-half-hsd}
\end{equation}
The above coupling steers the state of the system to the singlet only from the $\ket{T_0}$ state and leaves the $\ket{T_\pm}$ subspace untouched. 
However, the state can be \emph{shaken} out of the $\ket{T_\pm}$ subspace onto the $\ket{T_0}$ via a system-Hamiltonian of the form
\begin{equation}
	H_\s = \bm{\theta_1}\cdot\bm{\sigma_1}+\bm{\theta_2}\cdot\bm{\sigma_2},
	\label{eq:shake-ham-spin-half}
\end{equation}
where $\bm{\theta_{1/2}}$ can be arbitrary vectors \footnote{Exceptions are special choices of $\bm{\theta}$ for which $H_\s$ has an eigenstate which lives purely in the $\ket{T_\pm}$ subspace. Such a special choice does not transfer weight from that particular state in the $\ket{T_\pm}$ to the $\ket{T_0}$ and hence in the steady state, some weight remains in the orthogonal subspace.}.

The time evolution map of the density matrix of the system can be written explicitly as
\begin{subequations}
\begin{equation}
	\rho_\s\left(t+\frac{\delta t}{2}\right) = \mathrm{Tr}_\a[e^{-i \hsd\frac{\delta t}{2}}\rho_\a\otimes\rho_\s(t)e^{i \hsd\frac{\delta t}{2}}],
\end{equation}
\begin{equation}
	\rho_\s\left(t+\delta t\right) = e^{-i H_\s \frac{\delta t}{2}}\rho_\s\left(t+\frac{\delta t}{2}\right)e^{i H_\s \frac{\delta t}{2}},
\end{equation}
\label{eq:shake-and-steer-singlet-protocol}
\end{subequations}
with $\hsd$ and $H_\s$ given by Eqs.~\eqref{eq:shake-and-steer-spin-half-hsd} and \eqref{eq:shake-ham-spin-half} respectively.
Simulating the protocol described above again shows steering to pure singlet state: see Fig.~\ref{fig:shake-and-steer-singlet}(a). 
The Lindblad equation describing the dynamics in the time-continuum limit now also has a unitary part in addition to the dissipative part. The spectrum of the corresponding Lindbladian has a unique zero eigenvalue: see Fig.~\ref{fig:shake-and-steer-singlet}(b).

Such protocols can be easily generalised to the case of the AKLT state. For a pair of spins-1, a single system-detector coupling steers weight out of one of the $S^\mathrm{tot}=2$ states and a Hamiltonian rotates the system within the $S^\mathrm{tot}=2$ subspace in a sufficiently general fashion.

\section{Monotonicity of steering to the AKLT state \label{sec:aklt-monotonic}}
The numerical results in Sec.~\ref{sec:aklt} show that the AKLT ground state is the unique steady state of our measurement dynamics and steering to it is guaranteed via our protocol. However, the approach to the steady state is not necessarily monotonic and can depend on the distance measure used. In this section, we show explicitly that such a situation arises using $E_\aklt$ [Eq.~\eqref{eq:EAKLT}] as the distance measure. We then compare the results with those of other distance measures.

Since $H_\aklt = \sum_\ell\p{\ell}$ is a sum of projectors for each bond, we would like to understand the change in $\tr[\p{\ell^\prime}\rho_\s(t)]$ when the bond between site $\ell$ and $\ell+1$ is steered. In the following we will use the shorthand $\Pl = \p{\ell}$, $\Plp = \p{\ell^\prime}$, and $U_i=U_i^{(\ell,\ell+1)}$.

Using Eq.~\eqref{eq:multi-rho-evol}, it can be shown that
\begin{equation}
\begin{split}
{\rm Tr}[{\mathcal P}^\prime\rho_\s(t+\delta t)] =& {\rm Tr}[{\mathcal P}^\prime\rho_\s(t)]+{\rm Tr}\sum_n [U_n^\dagger {\mathcal P}^\prime U_n \rho(t)] \sin^2 \delta t\\
&- {\rm Tr}[({\mathcal P}{\mathcal P}^\prime +{\mathcal P}^\prime{\mathcal P})\rho_\s(t)](1-\cos \delta t)\\
&+ {\rm Tr}[{\mathcal P}{\mathcal P}^\prime{\mathcal P}\rho_\s(t)](1-\cos \delta t)^2.
\end{split}
\label{eq:full}
\end{equation}
The above equation has three different cases
\begin{itemize}
	\item $\ell=\ell^\prime$: In this case Eq.~\eqref{eq:full} reduces to 
	\begin{equation}\label{single}
{\rm Tr}[{\mathcal P}\rho_\s(t+\delta t)] = {\rm Tr}[{\mathcal P}\rho_\s(t)]  \cos^2 \delta t\,,
\end{equation}
which shows that the bond $(\ell,\ell+1)$ is steered towards the AKLT state.
\item $\ell$ and $\ell^\prime$ are such that the two bonds do not share a common site. In this case, ${\rm Tr}[{\mathcal P^\prime}\rho_\s(t+\delta t)] = {\rm Tr}[{\mathcal P^\prime}\rho_\s(t)]$ since the steering has no effect on the reduced density matrix of the bond $(\ell^\prime,\ell^\prime+1)$.
\item $\ell$ and $\ell^\prime$ are adjacent such that the two bonds in question share a common site. In this case
\begin{equation}
\begin{split}
{\rm Tr}[{\mathcal P}^\prime(\rho_\s(t+\delta t)-\rho_\s(t))] =& -2 \sin^2(\delta t/2) {\rm Tr} [{\mathcal A} \rho_\s(t)] \\
&+ 4 \sin^4(\delta t/2) {\rm Tr}[{\mathcal B} \rho_\s(t)],
\end{split}
\label{eq:prime}
\end{equation}
where
\begin{subequations}
\begin{equation}
{\mathcal A} = {\mathcal P}{\mathcal P}^\prime +{\mathcal P}^\prime{\mathcal P} - 2\sum_n U_n^\dagger {\mathcal P}^\prime U_n
\end{equation}
\begin{equation}
{\mathcal B} = {\mathcal P} {\mathcal P}^\prime {\mathcal P} - \sum_n U_n^\dagger {\mathcal P}^\prime U_n\,.
\end{equation}
\label{eq:opAB}
\end{subequations}
\end{itemize}

The change in energy of the system measured with respect to $H_\aklt$ can be obtained by summing Eq.~\eqref{eq:prime} over $\ell^\prime$ which gives
\begin{widetext}
\begin{equation}
\Delta E_\aklt(t) =  -2 \sin^2(\delta t/2) {\rm Tr} [(2{\mathcal P}+ {\mathcal A}_-+{\mathcal A}_+) \rho_\s(t)] + 4 \sin^4(\delta t/2) {\rm Tr}[({\mathcal P} + {\mathcal B}_-+{\mathcal B}_+) \rho_\s(t)]\,,
\end{equation}
\end{widetext}
where the operators $\mathcal{A}_\pm$ and $\mathcal{B}_\pm$ are defined in Eq.~\eqref{eq:opAB} with $\ell^\prime=\ell\pm 1$. To leading order in $\delta t$, a sufficient condition for the steering to be monotonic would be that four-site operator $\hat{\mathcal{E}}_\ell=\mathcal{P}+(\mathcal{A}_++\mathcal{A}_-)/2$ has only non-negative eigenvalues. However it turns out that the spectrum of $\hat{\mathcal{E}}$ has both positive and negative eigenvalues. Denoting the eigenvalues and eigenvectors of $\hat{\mathcal{E}}$ as $w$ and $\ket{w}$, the condition on $\rho_s(t)$ for $E_\aklt(t)$ to decay (grow) is given by $\mathrm{Tr}[\hat{\mathcal{E}}\rho_\s(t)]=\sum_w w\braket{w|\rho_\s(t)|w}\gtrless 0$.

\begin{figure}[t]
\includegraphics[width=\linewidth]{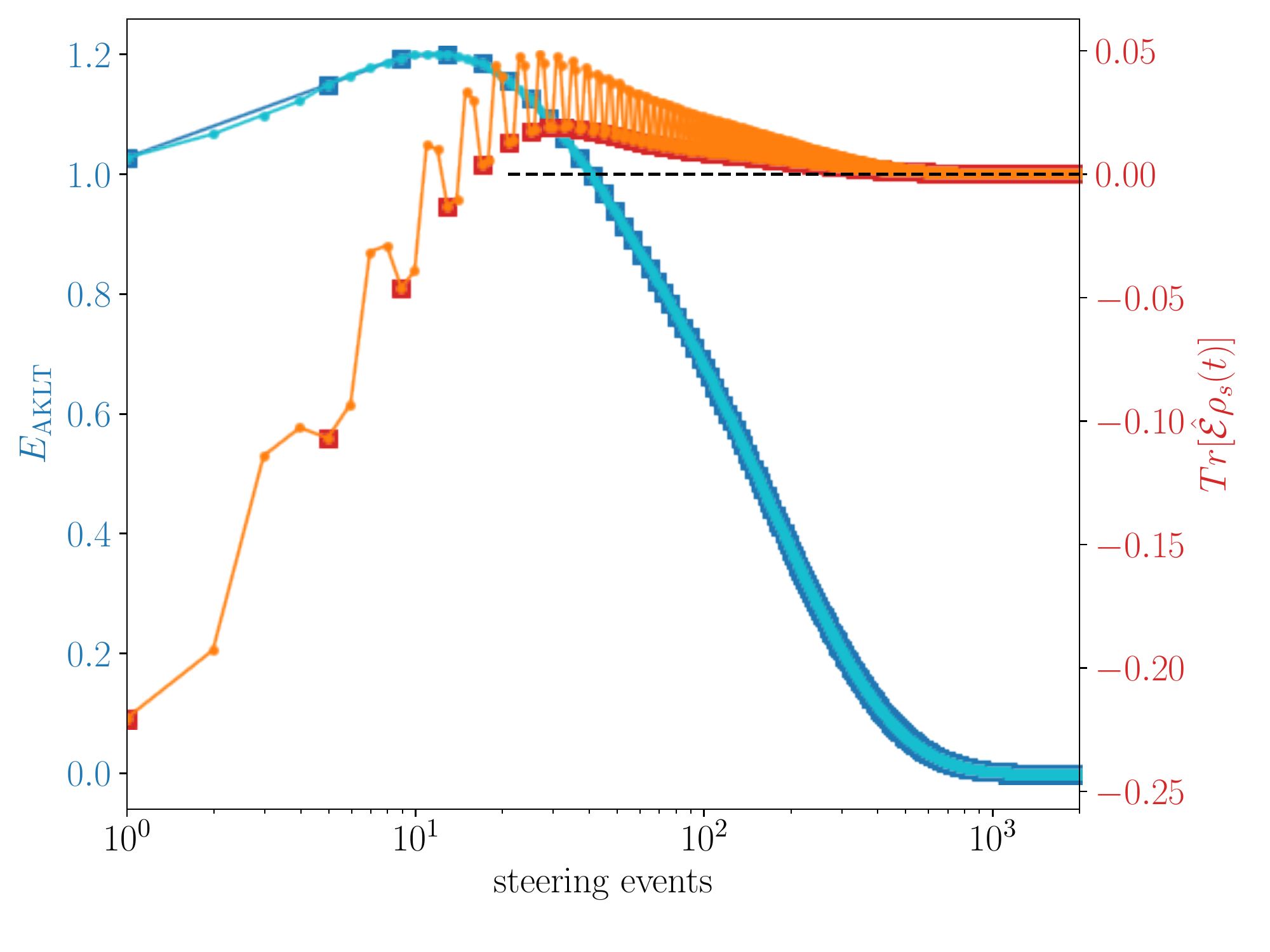}
\caption{Evolution of $\tr[\hat{\mathcal{E}}\rho_\s(t)]$ (red,orange) and $E_\aklt$ (blue,cyan) for an initial state which is the eigenstate corresponding to the most negative eigenvalue of $\mathcal{\hat{E}}$. Results are shown for a $N=4$ spin-1 chain. The dots correspond to data points after steering each bond whereas the squares correspond to data points after steering each bond of the system once.}
\label{fig:init}
\end{figure}

This observation allows us to specifically construct density matrices which show non-monotonic steering as measured by $E_\aklt$. For example, consider the initial density matrix as $\rho_\s(t=0)=\ket{w_-}\bra{w_-}$ where $w_-$ is the most negative eigenvalue of $\hat{\mathcal{E}}$. For such a state $\mathrm{Tr}[\hat{\mathcal{E}}\rho_\s(t)]$ is trivially negative. Numerical results for such a situation are shown in Fig.~\ref{fig:init} where $E_\aklt(t)$ initially grows while $\mathrm{Tr}[\hat{\mathcal{E}}\rho_\s(t)]<0$ and starts decaying only when $\mathrm{Tr}[\hat{\mathcal{E}}\rho_\s(t)]>0$.

\begin{figure}[t]
\includegraphics[width=\linewidth]{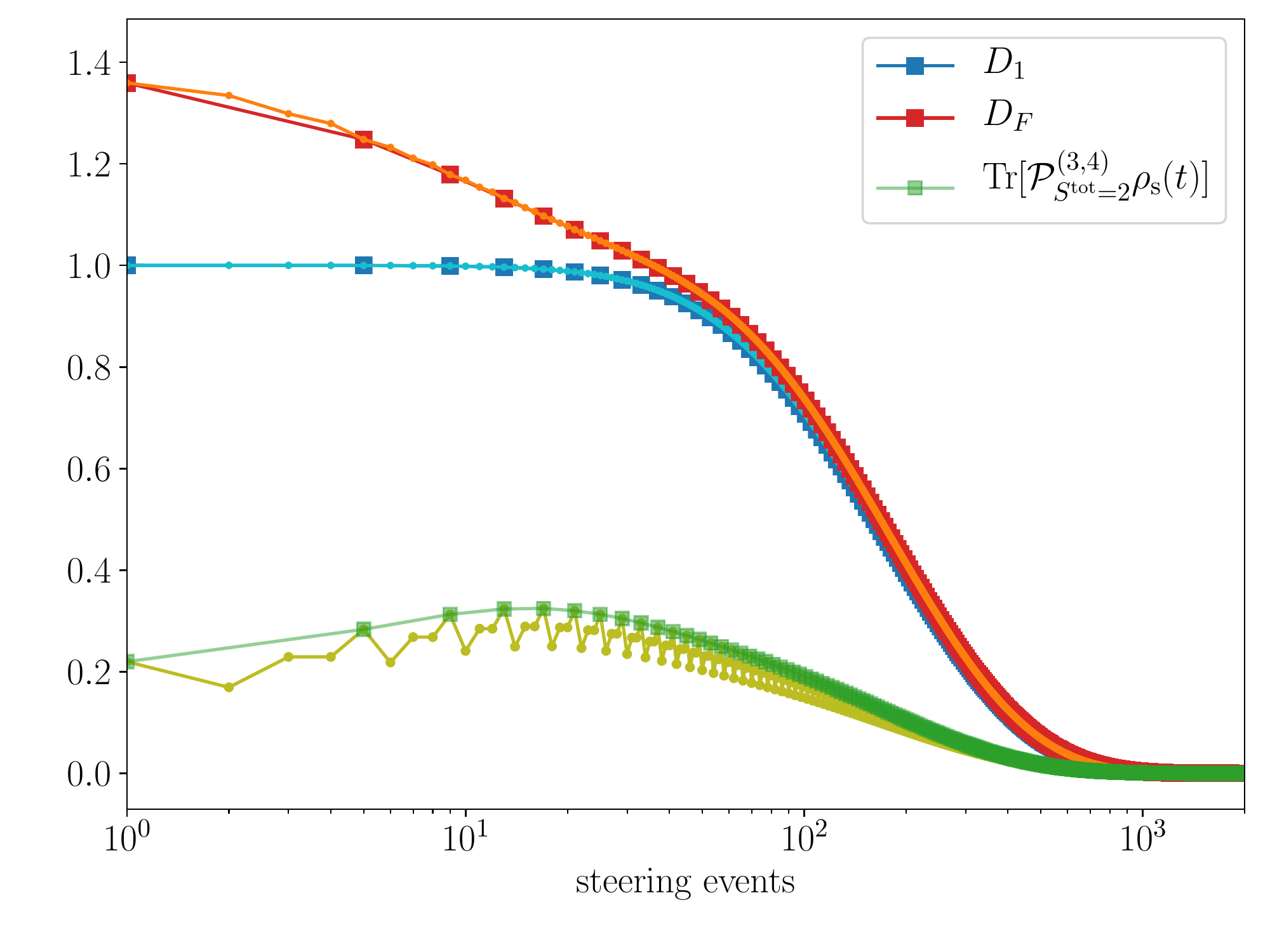}
\caption{Evolution of the global and local measures of distance from the AKLT ground state for the initial state used in Fig.~\ref{fig:init} (the eigenstate corresponding to the most negative eigenvalue of $\mathcal{\hat{E}}$). The Frobenius distance from the target state has a monotonic decay whereas the trace distance at early times does not change, although it does eventually decay to zero. The reduced density matrix of a particular bond also gets steered to that of the AKLT ground state but in a non-monotonic fashion.
Results are shown for a $N=4$ spin-1 chain. Similar to Fig.~\ref{fig:init} the dots correspond to data points after steering each bond whereas the squares correspond to data points after steering each bond of the system once.}
\label{fig:local-global}
\end{figure}

For such an initial state, it is also interesting compare the behaviour of other distance measures, both spatially local such as $\tr[\p{\ell}\rho_\s(t)]$ and global such as $D_1(t)$ and $D_F(t)$. The results are shown in Fig.~\ref{fig:local-global}. The distance of a local reduced density matrix measured via $\tr[\p{\ell}\rho_\s(t)]$ shows a non-monotonic behaviour. This is a manifestation of the phenomenon that when the bond adjacent to the one being monitored is steered, the former is steered away from the target as such the total energy grows. 

Turning to global measures of distance, the trace distance, $D_1(t)$, at early times is flat and it decays to zero only at late times. This is analogous to the situation described in the main text where the overlap of the state with the AKLT ground state stays zero but the state keeps evolving. The Frobenius distance, $D_F(t)$, on the other hand decays monotonically to zero.

The above results show that for many-body systems, monotonicity of steering and the exact condition for convergence to the target state can depend on the distance measure employed. Crucially, however, these results along with the numerical results in Sec.~\ref{sec:aklt} show that the system is guaranteed to be steered to the AKLT ground state.

\bibliography{refs}

\end{document}